\documentclass{elsart}

\usepackage{graphicx}

\usepackage{amssymb}

\begin{document}

\begin{frontmatter}



\title{Linear high-resolution  schemes for hyperbolic conservation laws: TVB numerical evidence\thanksref{support}}


\author[Palma]{C.~Bona\thanksref{Prague}},
\author[Palma]{C.~Bona-Casas\thanksref{fellow1}},
\author[Leuven]{J.~Terradas\thanksref{fellow2}}
\thanks[support]{Work jointly supported by the European Commission FEDER
funds, the Spanish Ministry of Science and the Balearic Islands
Government}
\thanks[Prague]{C.~Bona acknowledges the Charles University in Prague for
his hospitality during the completion of this work and specially
Tomas Ledvinka for useful suggestions and discussions.}
\thanks[fellow1]{C.~Bona-Casas acknowledges the support from the
FPU/2006-02226 fellowship of the Spanish Ministry of Science and
Education}
\thanks[fellow2]{J.~Terradas acknowledges the support from the Research
Council fellowship F/06/65 of the Katholieke Universiteit Leuven}

\address[Palma]{Departament de F\'{i}sica. Universitat de les Illes Balears.\\
Institute for Applied Computation with Community Code (IAC$\,^3$)}
\address[Leuven]{Centre for Plasma Astrophysics,  Katholieke Universiteit Leuven,
Celestijnenlaan 200B, B-3001 Leuven, Belgium}

\begin{abstract}
The Osher-Chakrabarthy family of linear flux-modification schemes
is considered. Improved lower bounds on the compression factors
are provided, which suggest the viability of using the unlimited
version. The LLF flux formula is combined with these schemes in
order to obtain efficient finite-difference algorithms. The
resulting schemes are applied to a battery of numerical tests,
going from advection and Burgers equations to Euler and MHD
equations, including the double Mach reflection and the
Orszag-Tang 2D vortex problem. Total-variation-bounded behavior is
evident in all cases, even with time-independent upper bounds. The
proposed schemes, however, do not deal properly with compound
shocks, arising from non-convex fluxes, as shown by
Buckley-Leverett test simulations.
\end{abstract}

\end{frontmatter}

\section{Introduction}

The study of hyperbolic conservation laws, as described by
\begin{equation}\label{Flux_Con}
    \partial_t u + \partial_x f(u) = 0~,
\end{equation}
is a classical topic in Computational Fluid Dynamics (CFD). We
have noted here by $u$ a generic array of dynamical fields, and we
will assume strong hyperbolicity, so that the characteristic
matrix
\begin{equation}\label{char_mat}
    A(u) = \partial f/\partial u
\end{equation}
has real eigenvalues and a full set of eigenvectors.

As it is well known, the system (\ref{Flux_Con}) admits weak
solutions, so that the components of $u$ may show piecewise-smooth
profiles. Standard finite-difference schemes, like the
Lax-Wendroff~\cite{Lax_Wendroff}  or MacCormack~\cite{MacCormack}
ones, produce spurious overshots and oscillations at non-smooth
points which can mask the physical solutions, even leading to code
crashing. These deviations do not diminish with resolution, in
analogy with the Gibbs phenomenon found in the Fourier series
development of discontinuous functions.

This difficulty was overcome in the pioneering work of
Godunov~\cite{Godunov}. On a uniform computational grid $x_j =
j\triangle x$, equation (\ref{Flux_Con}) can be approximated by
the semi-discrete equation
\begin{equation}\label{Flux_deriv}
    \partial_t u_j = - \frac{1}{\Delta x}~(h_{j+1/2}-h_{j-1/2})~,
\end{equation}
where the interface flux $h_{j+1/2}$ is computed by an
upwind-biased formula from the neighbor grid nodes. In the scalar
case, one can define the total variation of a discrete function as
\begin{equation}\label{TV}
    TV(u) = \sum_j~|\,u_j - u_{j-1}|~.
\end{equation}
In the case of systems, the total variation is defined as the sum
of the total variation of the components. Godunov scheme is
total-variation-diminishing (TVD), meaning that $TV(u)$ does not
increase during numerical evolution. It is obvious that TVD
schemes can not develop spurious oscillations: monotonic initial
data preserve their monotonicity during time evolution. Moreover,
the TVD property can be seen as a strong form of stability: any
blow-up of the numerical solution is excluded, as far as it would
increase the total variation.

Godunov scheme is the prototype of the so-called upwind-biased
schemes, which require either the exact or some approximate form
of spectral decomposition of the characteristic matrix
(\ref{char_mat}). This makes them both computationally expensive
and difficult to extend to the multidimensional case. A much
simpler alternative is provided by the local Lax-Friedrichs (LLF)
scheme (Rusanov scheme~\cite{Rusanov})
\begin{equation}\label{Lax_Fried}
    h_{j+1/2} = \frac{1}{2}~[f_{j+1} + f_j
    - \lambda_{j+1/2}\, (u_{j+1} - u_j)\,]~,
\end{equation}
where $\lambda$ is the spectral radius of the characteristic
matrix and we have taken
\begin{equation}\label{lambda_nonlocal}
    \lambda_{j+1/2} = max(\lambda_j,\lambda_{j+1})~.
\end{equation}

The LLF scheme is the prototype of the so-called centered schemes,
which deserve a revived interest nowadays in view of
multidimensional applications. It clear from (\ref{Flux_deriv},
\ref{Lax_Fried}) that the LLF scheme, like the Godunov one, is
only first-order accurate in space. Second-order accuracy can be
obtained following the Harten modified-flux
approach~\cite{Harten83}, which was soon extended to very-high
accuracy (up to 15th order) by Osher and
Chakrabarthy~\cite{ICASE}. The basic idea is to replace the lower
order TVD flux $h_{j+1/2}$ by a modified flux $f_{j+1/2}$,
obtained by some interpolation procedure involving a higher number
of nodes.

All these high-resolution schemes require some form of
flux-correction limiters in order to ensure the TVD property. As a
consequence, accuracy is reduced to (at most) first order at
non-sonic critical points, where the limiters come into play. In
order to circumvent this problem, one can relax the TVD condition,
demanding instead that the total variation is bounded, that is
\begin{equation}\label{TVB}
    TV(u) \leq B~,
\end{equation}
where the upper bound $B$ is independent of the resolution, but
could depend on the elapsed time. Even if we are ready to relax
the stronger TVD requirement, keeping the bound (\ref{TVB}) is
important from the theoretical point of view. One major advantage
of total-variation-bounded (TVB) schemes is that there is a
convergent (in $L^1_{loc}$) subsequence as $\Delta x \rightarrow
0$ to a weak solution of (\ref{Flux_Con}). If an additional
entropy condition is satisfied, then the proposed scheme is
convergent (see for instance Ref.~\cite{Leveque}).

An interesting example of such TVB schemes was given by
Shu~\cite{Shu87}, by softening the flux limiters proposed in
Ref.~\cite{ICASE}. Although the TVB property is proven for the
schemes presented in~\cite{Shu87}, based on a linear
flux-modification procedure, a rigorous proof is still unavailable
for more complex cases. An important example is provided by the
essentially-non-oscillatory (ENO) methods~\cite{ENOa}~\cite{ENOb},
where the TVD property is relaxed in a different way. Numerical
evidence shows that ENO schemes, as well as their weighted-ENO
variants~\cite{Balsara-Shu}-\cite{BL06}, deserve their name: the
TVB property is satisfied in practice, even with time-independent
bounds. An implementation of these high-resolution methods for the
LLF Flux is given in Refs.~\cite{KT00}~\cite{KL00}.

In this paper we consider (the unlimited version of) the
'$\beta$-family' of Osher-Chakrabarthy linear flux-modification
algorithms in the semidiscrete framework. The choices of the
$\beta$ parameter are optimized for their use with the unlimited
version, by refining the bounds given in the original
paper~\cite{ICASE}. As a lower order TVD flux, we will use the
standard LLF one (\ref{Lax_Fried}). The resulting high-resolution
schemes can then be recast into a compact finite-difference
formula. This greatly increases computational efficiency,
specially with a view to three-dimensional applications. We
perform a battery of standard tests in one space dimension,
covering advection, Burgers and Euler equations, in order to show
that the TVB property is fulfilled in practice for the selected
values of the $\beta$ parameter. We are not able, however, of
getting the right result for compound shocks, arising from
non-convex fluxes; this is illustrated by the Buckley-Leverett
test simulations. We also consider some multidimensional tests
cases with the Euler and magneto-hydrodynamics (MHD) equations,
including the double Mach reflection and the Orszag-Tang 2D vortex
problem. Total-variation-bounded behavior is evident in all the
proposed cases, even with time-independent upper bounds.

\section{The Osher-Chakravarthy $\beta$-schemes}
Following Ref.~\cite{ICASE}, let us consider the centered $2m-1$
order schemes:
\begin{equation}\label{Dform_gen}
        \partial_t u_j = - C^{2m}f_j + (-1)^{m-1}\beta (\Delta x)^{2m-2}
        D_{+}^{m}D_{-}^{m-1}(df^{+}_{j-1/2}-df^{-}_{j-1/2})~,
\end{equation}
where $C^{2m}$ is the central $2m$th-order-accurate difference
operator with a stencil of $2m+1$ grid points, and we have used
the standard notation $D_{\pm}$ for the elementary difference
operators. The flux differences $df^{\pm}$ are defined as follows:
\begin{equation}\label{dfpm}
    df^{+}_{j+1/2} = f_{j+1} - h_{j+1/2}\qquad
    df^{-}_{j+1/2} = h_{j+1/2} - f_{j}~,
\end{equation}
where $h_{j+1/2}$ is any (lowest-order) TVD flux. The $\beta$
parameter in the dissipative term in (\ref{Dform_gen}) is assumed
to be positive: a necessary condition for stability.

The algorithms (\ref{Dform_gen}) can be put into an explicit
flux-conservative form by replacing the lowest order flux
$h_{j+1/2}$ in (\ref{Flux_deriv}) by~\cite{ICASE}~\cite{Shu87}
\begin{equation}\label{flux_develop}
    f_{j+1/2} = h_{j+1/2} + \sum_{k=-m+1}^{m-1}
    \left( c^{m}_{~k}~df^{-}_{j+k+1/2}
    +d^{m}_{~k} ~df^{+}_{j+k+1/2} \right)~,
\end{equation}
where
\begin{equation} \label{dcdefs}
    d^{m}_{~k} = \nu^m_{~k} - (-1)^k\beta \left(
    \begin{array}{c}
      2m-2 \\
      k+m-1 \\
    \end{array} \right)~, \qquad
    c^{m}_{~k} = -d^{m}_{~-k}
\end{equation}
(we use here a compact notation), and
\begin{eqnarray}
  \nu_0^m &=& 1/2~, \qquad\qquad \nu^m_k = -\nu^m_{~-k} \qquad (k \ne 0)\\
  \nu_{m-1}^m  &=& (-1)^{m -1} \left[ m \left(\begin{array}{c}
    2m \\
    m \\
  \end{array}\right)\right]^{-1} \qquad (m > 1)\\
    \nu_k^{m+1}  &=& \nu_k^m + (-1)^{k}~\frac{k}{m}~\left(\begin{array}{c}
    2m \\
    m-k \\
  \end{array}\right)\left[ (m+1) \left(\begin{array}{c}
    2m+2 \\
    m+1 \\
  \end{array}\right)\right]^{-1}.
\end{eqnarray}

The TVD property is enforced by limiting the flux differences
$df^{\pm}$  (see~\cite{ICASE}, ~\cite{Shu87} for the details). As
a generic example, $df^{+}_{j+k+1/2}$ in (\ref{flux_develop}) is
replaced by
\begin{equation}\label{limiter}
    minmod(~df^{+}_{j+k+1/2}, bdf^{+}_{j+1/2},bdf^{+}_{j-1/2}~),
\end{equation}
where $b$ is a compression factor. This replacement introduces a
non-linear component in the linear flux-correction formula
(\ref{flux_develop}). The resulting scheme will be TVD if and only
if:
\begin{eqnarray}\label{positivitya}
  C_{j+1/2} \equiv 1 + \sum_{k=-m+1}^{m-1}
    c^{~m}_k ~\frac{df^{-}_{j+k+1/2}-df^{-}_{j+k-1/2}}{df^{-}_{j+1/2}} ~&\geq&~ 0 \\
    \label{positivityb}
  D_{j-1/2} \equiv 1 + \sum_{k=-m+1}^{m-1}
    d^{~m}_k ~\frac{df^{+}_{j+k+1/2}-df^{+}_{j+k-1/2}}{df^{+}_{j-1/2}} ~&\geq&~
    0~. \\ \label{positivityc}
    \lambda_j~\frac{\Delta t}{\Delta x}~(C_{j+1/2}+D_{j+1/2})
    ~&\le&~ 1~,
\end{eqnarray}
where we have assumed a time discretization based on the forward
Euler step, so that the last condition provides an upper bound on
the time step $\Delta t$.

\section{Compression factor optimization}
In the original paper~\cite{ICASE}, the ansatz
\begin{equation}\label{ansatz}
    \beta \le [~m \left(%
\begin{array}{c}
  2m \\
  m \\
\end{array}%
\right)  ~]^{-1}
\end{equation}
was used for getting a sufficient condition from
(\ref{positivitya}, \ref{positivityb}), amounting to a simple
constraint on the range of the compression parameter $b$
\begin{equation}\label{brange}
    0 ~<~ b ~\le~ [~1+2\beta\left(%
\begin{array}{c}
  2m-2 \\
  m-1 \\
\end{array}%
\right)~]~[~\sum_{j=2}^m ~\frac{1}{2j-1}~]^{-1}~.
\end{equation}

Allowing for (\ref{brange}), the upper bound $b_{max}$ increases
with $\beta$, which is in turn bounded by (\ref{ansatz}). For the
third-order scheme ($m=2$), the optimal choice would then be
$\beta=1/12$, so that the compression parameter may reach
$b_{max}=4$, still preserving the TVD property. This means that,
for monotonic profiles, the flux-correction limiters would act
only where the higher order corrections in neighboring
computational cells differ at least by a factor of four. This is
not to be expected in practical, good resolution, simulations of
smooth profiles, even when large gradients appear. This
high-compression-factor property can be at the origin of the
robust behavior of these schemes, even in their unlimited form, as
we will see in the numerical applications presented below.

As far as we are proposing to use the unlimited version, it makes
sense to find the choices of $\beta$ that maximize the compression
factor, going beyond the ansatz (\ref{ansatz}). Higher values of
$b_{max}$ can be actually obtained by a detailed case-by-case
study of the original TVD conditions (\ref{positivitya},
\ref{positivityb}). For instance, by reordering the terms in
(\ref{positivityb}) we get
\begin{equation}\label{positivityb2}
      D_{j-1/2} \equiv 1 + \sum_{k=-m+1}^{m}
    (d^{~m}_{k-1}-d^{~m}_{k})~\frac{df^{+}_{j+k-1/2}}{df^{+}_{j-1/2}} ~\geq~
    0~,
\end{equation}
where we assume $d^{~m}_k=0$ when $|k| \ge m$. A sufficient
condition for (\ref{positivityb2}) to hold is
\begin{equation}\label{newbound}
    1 + d^{~m}_{-1}-d^{~m}_{0} + b~\sum_{k\neq 0}
    min(d^{~m}_{k-1}-d^{~m}_{k},0) ~\geq~0~,
\end{equation}
which actually refines the former condition (\ref{brange}). The
same reasoning shows that, allowing for (\ref{dcdefs}), a
sufficient condition for (\ref{positivityc}) to hold is:
\begin{equation}\label{newcourbound}
       \lambda_j~\frac{\Delta t}{\Delta x}~[~d^{~m}_{-1}-d^{~m}_{0}
       +b~\sum_{k\neq 0}
    max(d^{~m}_{k-1}-d^{~m}_{k},0)~] ~\le~ 1/2~.
\end{equation}

For the simpler non-trivial cases we have (decreasing $k$
order):\begin{eqnarray}
  d^{~2}_k = (~\beta-\frac{1}{12},~&\frac{1}{2}-2\beta,&~\beta+\frac{1}{12}~) \\
  d^{~3}_k = (~\frac{1}{60}-\beta,~4\beta-\frac{7}{60},~&\frac{1}{2}-6\beta,&~
  4\beta+\frac{7}{60},~-\frac{1}{60}-\beta~)~.
\end{eqnarray}
For $m=2$, condition (\ref{newbound}) leads then to:
\begin{eqnarray}
  b &~\le~& 7/2 + 18\beta \qquad  (\beta \le \frac{1}{12})\\
  b &~\le~& \frac{7+36\beta}{1+12\beta} \qquad  (\frac{1}{12} \le \beta \le \frac{7}{36})~.
\end{eqnarray}
It follows that the optimal values for the third-order scheme are
\begin{equation}\label{optimal3}
    \beta = \frac{1}{12}~, \qquad b_{max} = 5~.
\end{equation}
For the fifth-order scheme ($m=3$), condition (\ref{newbound})
leads instead to:
\begin{eqnarray}
  b &~\le~& \frac{37+600\beta}{16} \qquad  (\beta \le \frac{1}{60})\\
  b &~\le~& \frac{37+600\beta}{15+60\beta} \qquad  (\frac{1}{60} \le \beta \le
  \frac{2}{75}) \\
  b &~\le~& \frac{37+600\beta}{7+360\beta} \qquad  (\frac{2}{75} \le \beta \le \frac{37}{600})~.
\end{eqnarray}
It follows that the optimal values for the fifth-order scheme are
\begin{equation}\label{optimal5}
    \beta = \frac{2}{75}~, \qquad b_{max} = \frac{265}{83}~.
\end{equation}
Note that the ansatz (\ref{ansatz}) gives a smaller compression
factor $b_{max} = 9/4$ and, more important, the optimal $\beta$
value in this case is beyond the original bound $1/60$. Note also
that the values of the compression parameter tend to diminish with
the accuracy order of the algorithm. This suggests that
higher-order cases $m>3$ may not be so useful in the unlimited
case.

\section{Finite difference version}
The linear flux-modification scheme described in the preceding
section can be applied to any lower-order TVD flux. The case of
the LLF flux (\ref{Lax_Fried}) has actually been considered
in~\cite{Shu87}. Our objective here is to obtain a scheme which
can be cast as a simple finite-difference algorithm, so that we
will take advantage of the simplicity of the LLF flux
(\ref{Lax_Fried}), which can be written in flux-vector-splitting
(FVS) form:
\begin{equation}\label{hsplit}
    h_{j+1/2} = f^+_j + f^-_{j+1}~, \qquad f^\pm_j \equiv
    \frac{1}{2}~[f_{j} \pm \lambda_{j\pm 1/2}\,u_j]~.
\end{equation}

The FVS form (\ref{hsplit}), like the original one
(\ref{Lax_Fried}), is just first-order accurate. We will extend it
to higher-order accuracy by means of the Osher-Chakrabarthy
algorithm, as described in the previous sections. The flux
differences (\ref{dfpm}) in this case get the simple form:
\begin{equation}\label{dfpm_new}
    df^{\pm}_{j+1/2} = 1/2~[\,f_{j+1}-f_{j}\,\pm\,\lambda_{j+1/2}\,
    (u_{j+1}-u_{j})\,]~.
\end{equation}

The linear character of this formula allows to get a compact
finite-difference expression for the whole scheme. Allowing for
(\ref{dfpm_new}), the semi-discrete algorithm (\ref{Dform_gen})
can be written as
\begin{equation}\label{Dform_new}
    \partial_t u_j = - C^{2m}f_j + (-1)^{m-1}\beta (\Delta x)^{2m-1}
        D_{+}^{m}D_{-}^{m-1}(\lambda_{j-1/2}\,D_{-}u_j)~,
\end{equation}
which amounts to assume a $2m$th-order-accurate central difference
operator acting on the flux terms plus a dissipation operator of
order $2m$ depending on the spectral radius $\lambda$. As we will
see below, the resulting finite-difference scheme
(\ref{Dform_new}) provides a cost-effective alternative for CFD
simulations.

Let us remark here that the choices (\ref{optimal3},
\ref{optimal5}) derived in the previous section are optimal for a
generic choice of the lowest-order TVD Flux. In the LLF case
(\ref{Lax_Fried}), however, it is clear that the spectral radius
can be multiplied by a global magnifying factor $K>1$, while
keeping the TVD properties. Allowing for the finite-difference
form (\ref{Dform_new}) of the unlimited version, magnifying
$\lambda$ amounts to magnify $\beta$, that is:
\begin{equation}\label{rescaling}
    (\beta,~K\lambda)\qquad\Leftrightarrow
    \qquad(K\beta,~\lambda)~.
\end{equation}
It follows that the values of the compression factor $b_{max}$
obtained in the previous section must be interpreted just as
lower-bound estimates. In particular, the equivalence
(\ref{rescaling}) implies that any compression factor bound
obtained for a particular value $\beta_0$ applies as well to all
values $\beta>\beta_0$. This agrees with the interpretation of the
second term in (\ref{Dform_new}) as modelling numerical
dissipation. On the other side, this dissipation term is actually
introducing the main truncation error. We will use then in what
follows the $\beta$ values in (\ref{optimal3}, \ref{optimal5}),
which are still optimal in the sense that they provide the lower
numerical error compatible with the highest lower-bound for the
compression parameter.

\section{One-dimensional numerical tests}\label{1D}
We will test now the behavior of the centered finite-difference
scheme (\ref{Dform_new}) by means of some standard numerical
experiments in one space dimension. We will use here the
well-known method-of-lines (MoL)~\cite{MoL} in order to deal
separately with the space and the time discretization. In every
case, the time discretization will be implemented by the following
strong-stability-preserving (SSP), 3rd-order-accurate, Runge-Kutta
algorithm~\cite{SSP}:
\begin{eqnarray}\label{RK3}\nonumber
    u^* &=& E(u^n) \\
    u^{**} &=& \frac{3}{4}~u^n + \frac{1}{4}~E(u^*)\\
    u^{n+1} &=& \frac{1}{3}~u^n + \frac{2}{3}~E(u^{**})~, \nonumber
\end{eqnarray}
where $E(u)$ is the basic Euler step, that is
\begin{equation}\label{Euler}
    E(u_j) = u_j + \Delta t~(\partial_t\,u_j)\,,
\end{equation}
and $\partial_t\,u_j$ is computed by the finite difference formula
(\ref{Dform_new}).

\subsection{Advection equation} Let us start by the scalar
advection equation. This is the simplest linear case, but it
allows to test the propagation of arbitrary initial profiles,
containing jump discontinuities and corner points, departing from
smoothness in many different ways. This is the case of the
Balsara-Shu profile~\cite{Balsara-Shu}, which will be evolved with
periodic boundary conditions.

\begin{figure}[t]
\centering
\includegraphics[width=12cm, height=4cm]{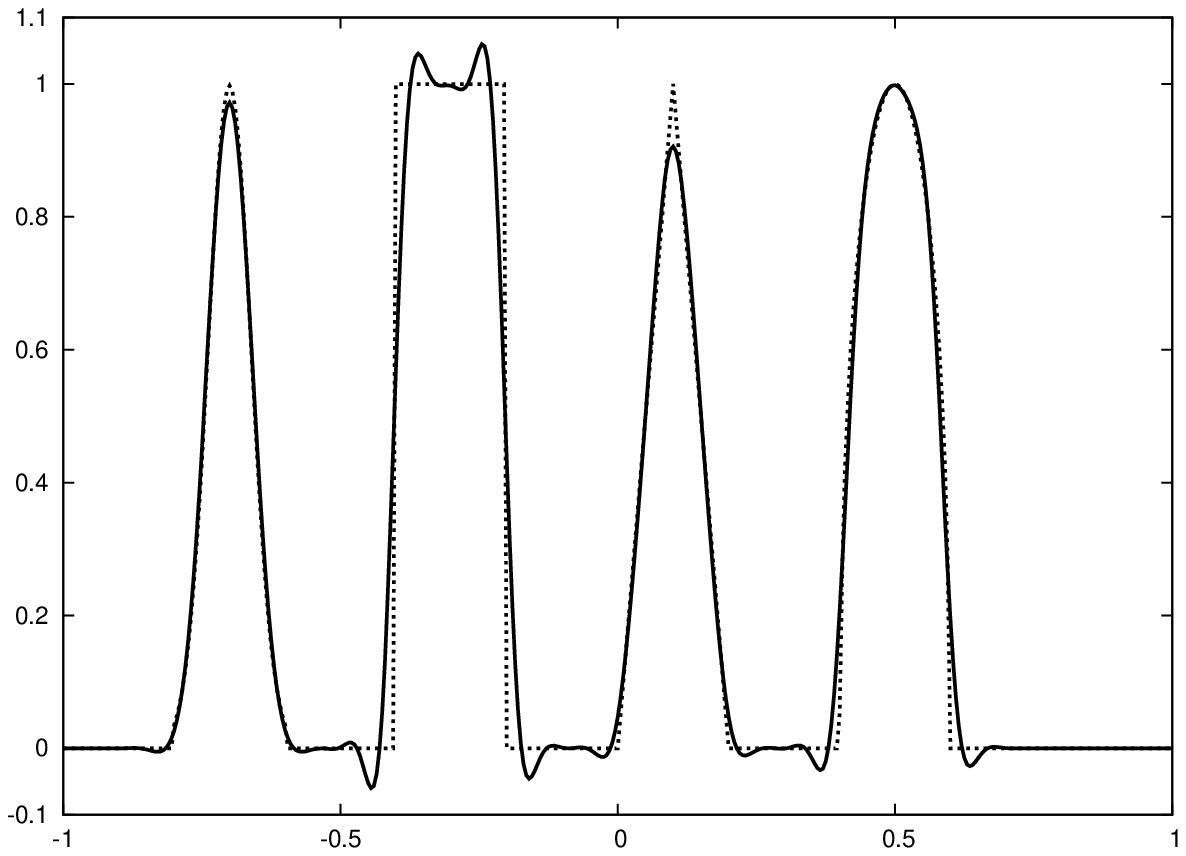}
\includegraphics[width=12cm, height=4cm]{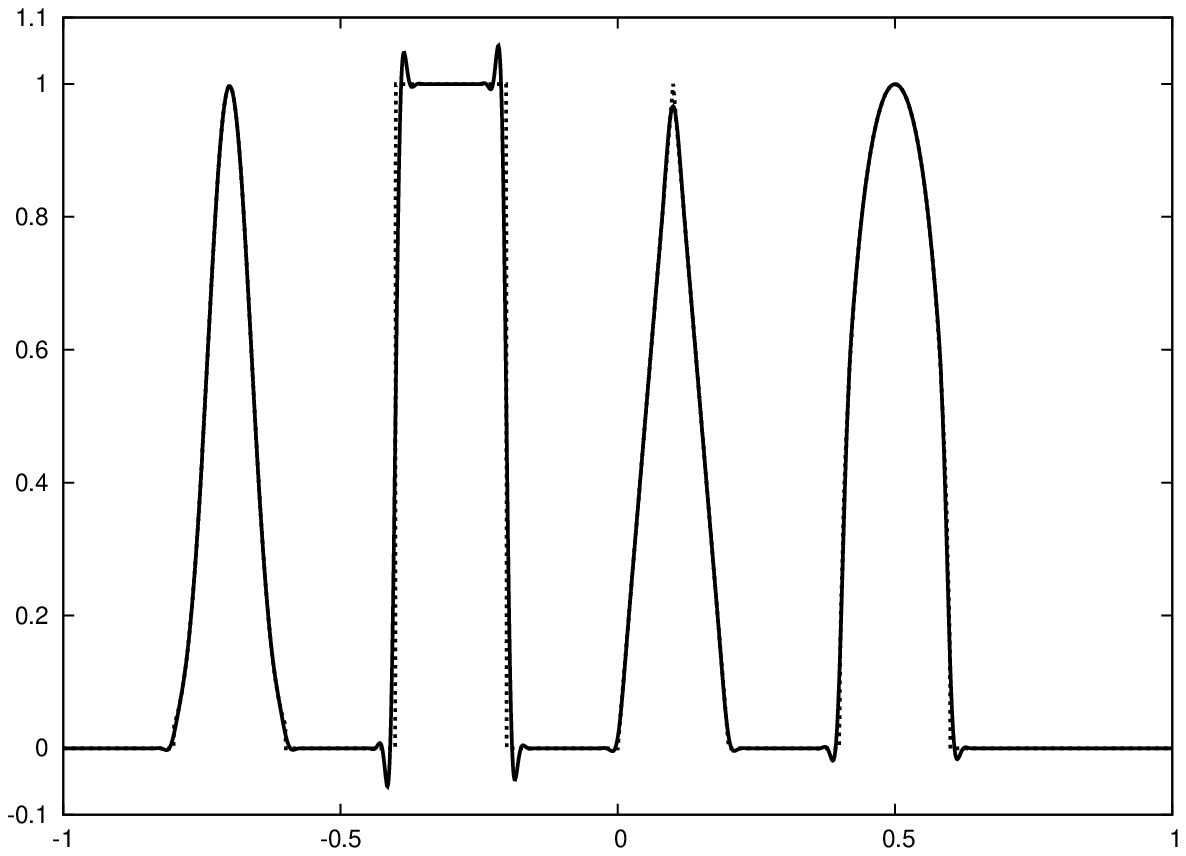}
\caption{Advection of the Balsara-Shu profile in a numerical mesh
of either $400$ points (upper panel) or $800$ points (lower
panel). A third-order scheme ($m=2$, $\beta=1/12$) is used in both
cases. The results are compared with the initial profile (dotted
line) after a single round-trip.} \label{adv1}
\end{figure}

We compare in Fig.~\ref{adv1} the numerical result with the exact
solution after a single round trip, for two different resolutions.
The third-order five-points formula from the proposed class
(\ref{Dform_new}) has been used with $\beta=1/12$ in both cases.
The propagation speed in the simulation agrees with the exact one,
as expected for a third-order-accurate algorithm. The smooth
regions are described correctly: even the height of the two
regular maxima is not reduced too much by dissipation, as expected
for an unlimited algorithm with just fourth-order dissipation.
There is a slight smearing of the jump slopes, as usual for
contact discontinuities, which gets smaller with higher
resolution.

\begin{figure}[b]
\centering
\includegraphics[width=12cm, height=4cm]{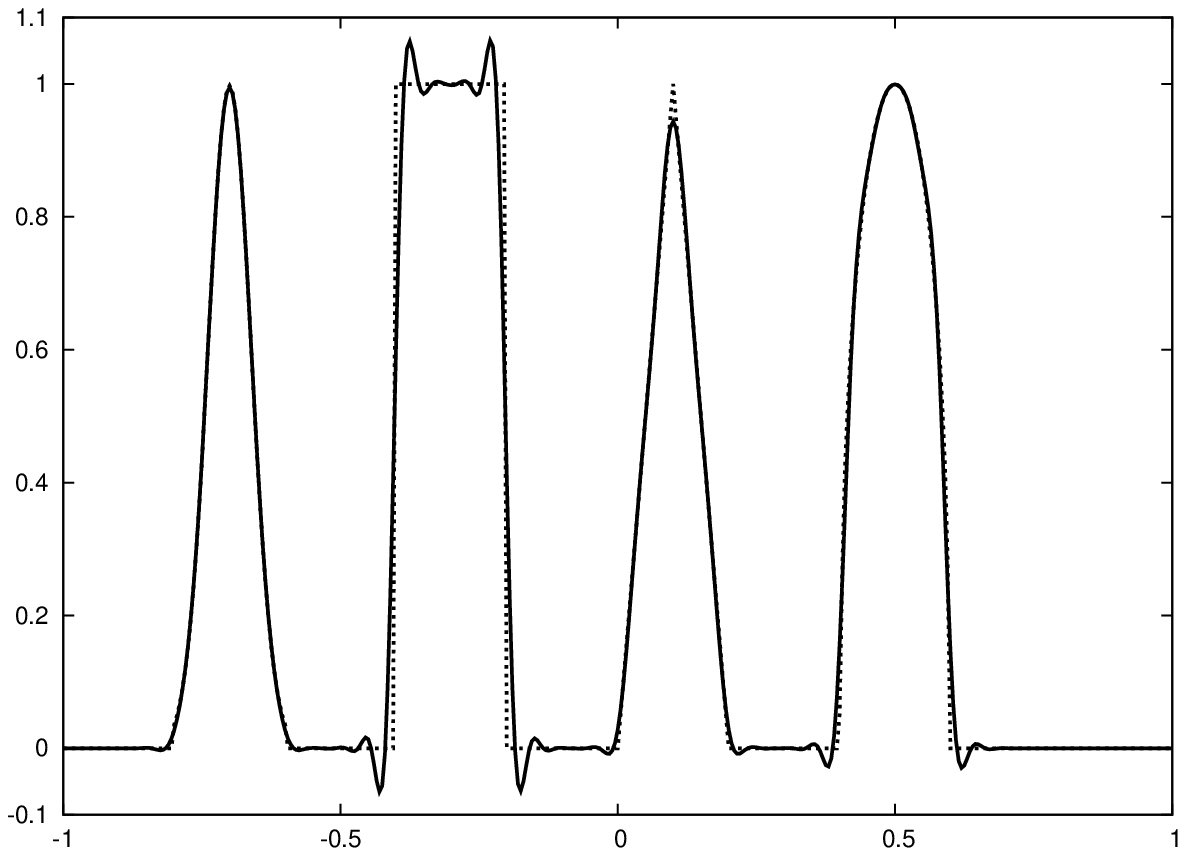}
\includegraphics[width=12cm, height=4cm]{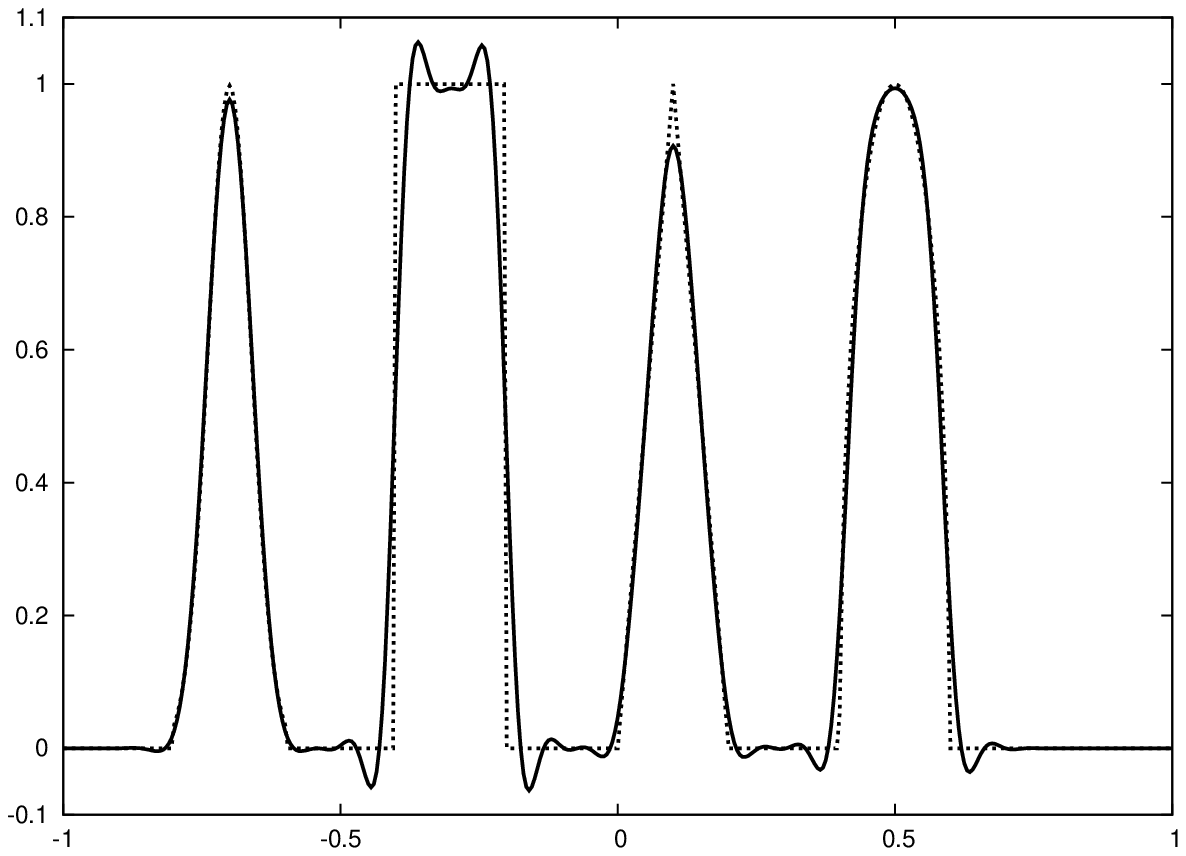}
\caption{Same as in Fig.~\ref{adv1}, but using a fifth-order
scheme ($m=3$) with $\beta=2/75$ (upper panel). In the lower panel
we show the results after ten round-trips. The same settings are
used in both cases.} \label{adv2}
\end{figure}

Concerning monotonicity, it is clear that the total variation of
the initial profile has increased by the riddles besides the
corner points and, more visibly, near the jump discontinuities. By
comparing the two resolutions, we see that the height of the
overshots does not change. This means that, as in the case of the
Gibbs phenomenon, there is no convergence by the maximum norm,
although convergence by the $L_2$ or similar norms is apparent
from the results.  On the other hand, it is clear that the total
variation is bounded for this fixed time, independently of the
space resolution or, equivalently, the time step size. This is
precisely the requirement for TVB.

\begin{figure}[t]
\centering
\includegraphics[width=12cm, height=5cm]{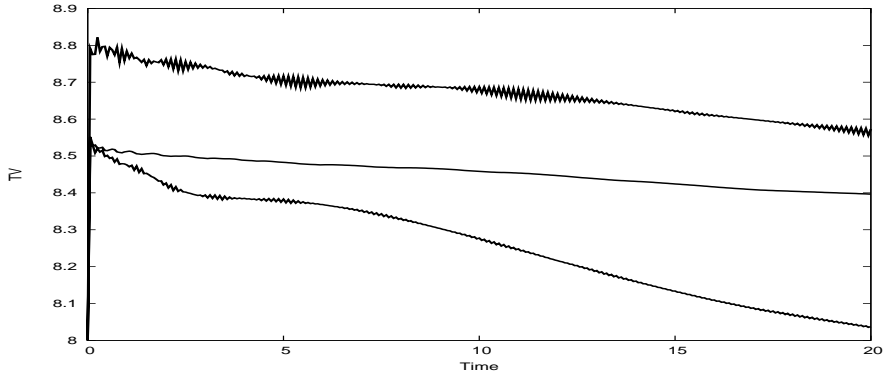}
\caption{Advection equation. Time evolution of the total
variation. The horizontal axis corresponds to the exact solution:
$TV(u)=8$. From top to bottom: $m=3$ scheme with 400 points, $m=2$
scheme with 800 points, and $m=2$ scheme with 400 points. After
the initial increase, which depends on the selected method, the TV
tends to diminish. Increasing resolution just reduces the TV
diminishing rate.} \label{tvevol}
\end{figure}

We show in Fig.~\ref{adv2} the same simulation, in a $400$ points
mesh, for the fifth-order method ($m=3$, $\beta=2/75$). In the
upper panel, corresponding to a single round-trip, we can see that
one additional riddle appears at every side of the critical
points, due to the larger (seven point) stencil. We show also in
the lower panel the results of the same simulation after ten
round-trips. The cumulative effect of numerical dissipation is
clearly visible: the extra riddles tend to diminish. The total
variation is not higher than the one after a single round trip.
This statement can be verified by plotting, as we do in
Fig.~\ref{tvevol}, the time evolution of $TV(u)$ for the different
cases considered here. In all cases, a sudden initial increase is
followed by a clear diminishing pattern. These numerical results
indicate that the bound on the total variation is actually
time-independent, beyond the weaker TVB requirement.

\subsection{Burgers equation}
\begin{figure}[t]
\centering
\includegraphics[width=0.49\textwidth]{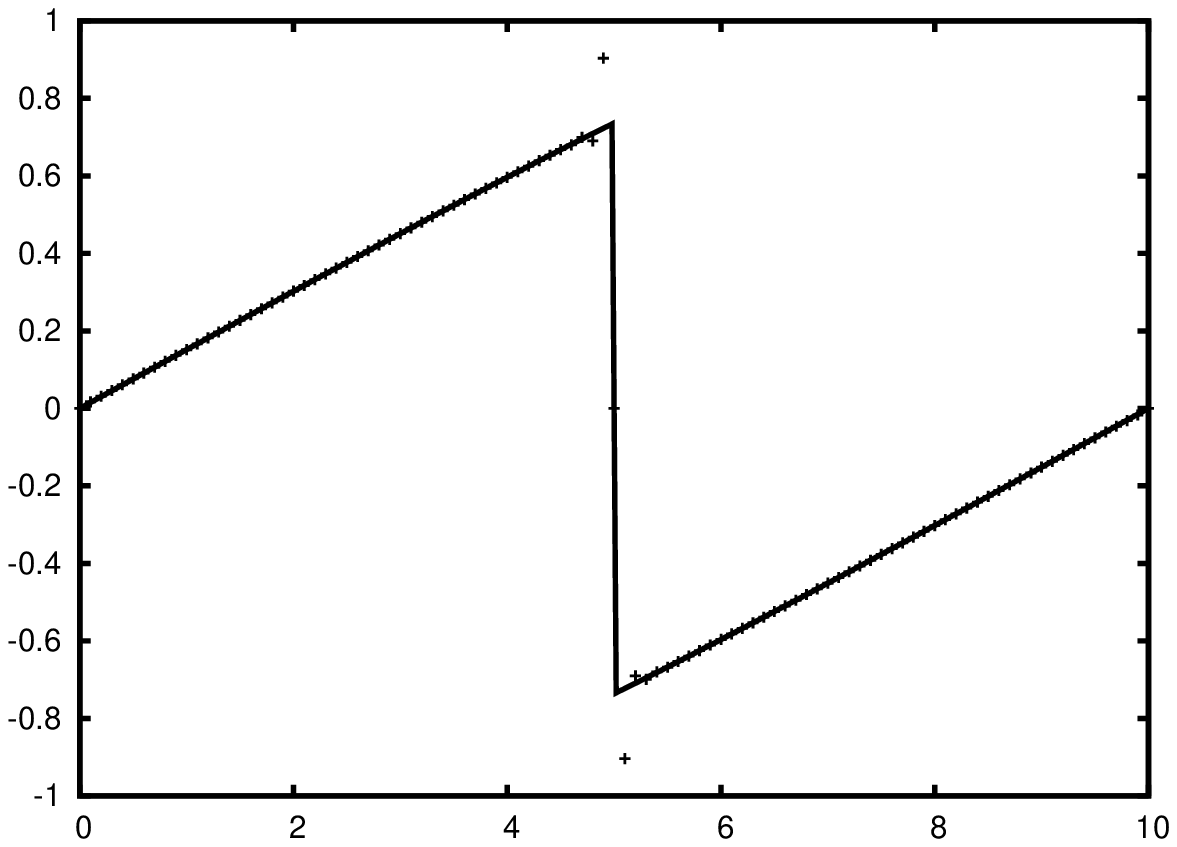}
\includegraphics[width=0.49\textwidth]{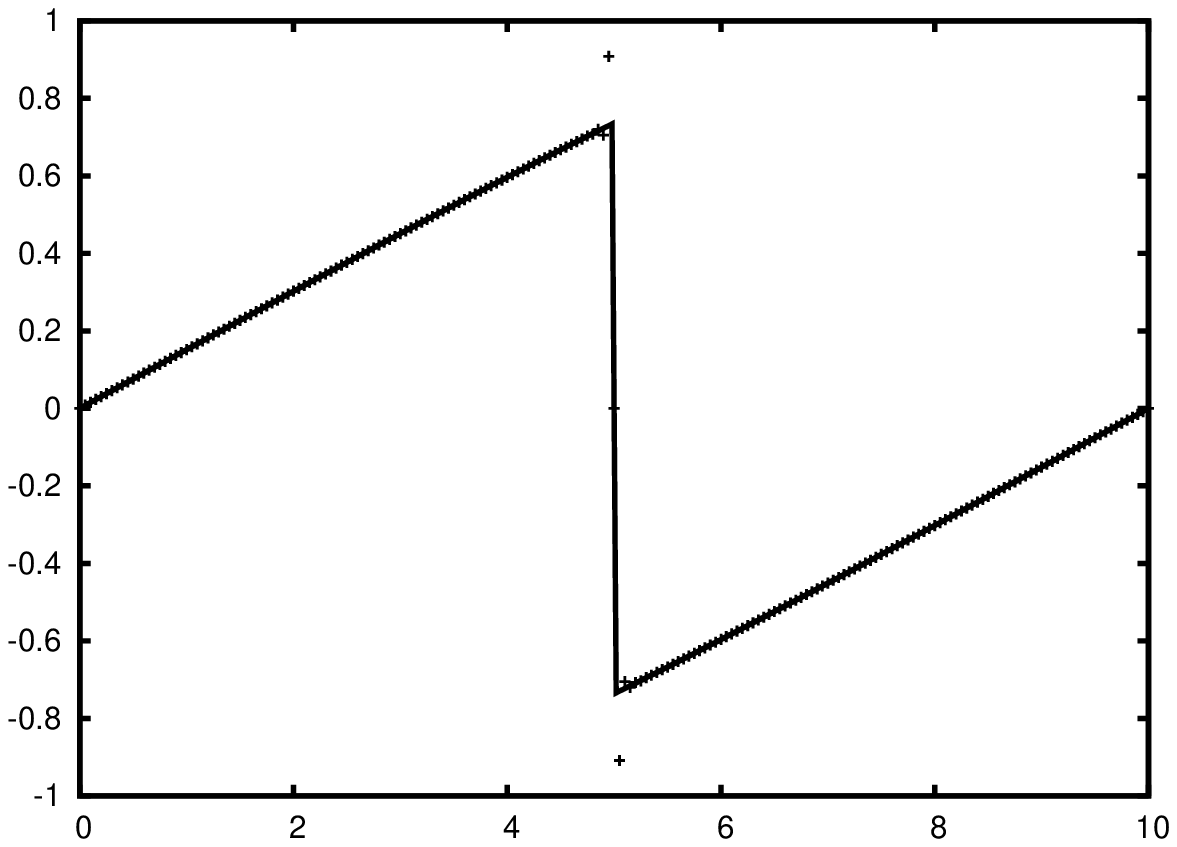}
\includegraphics[width=0.49\textwidth]{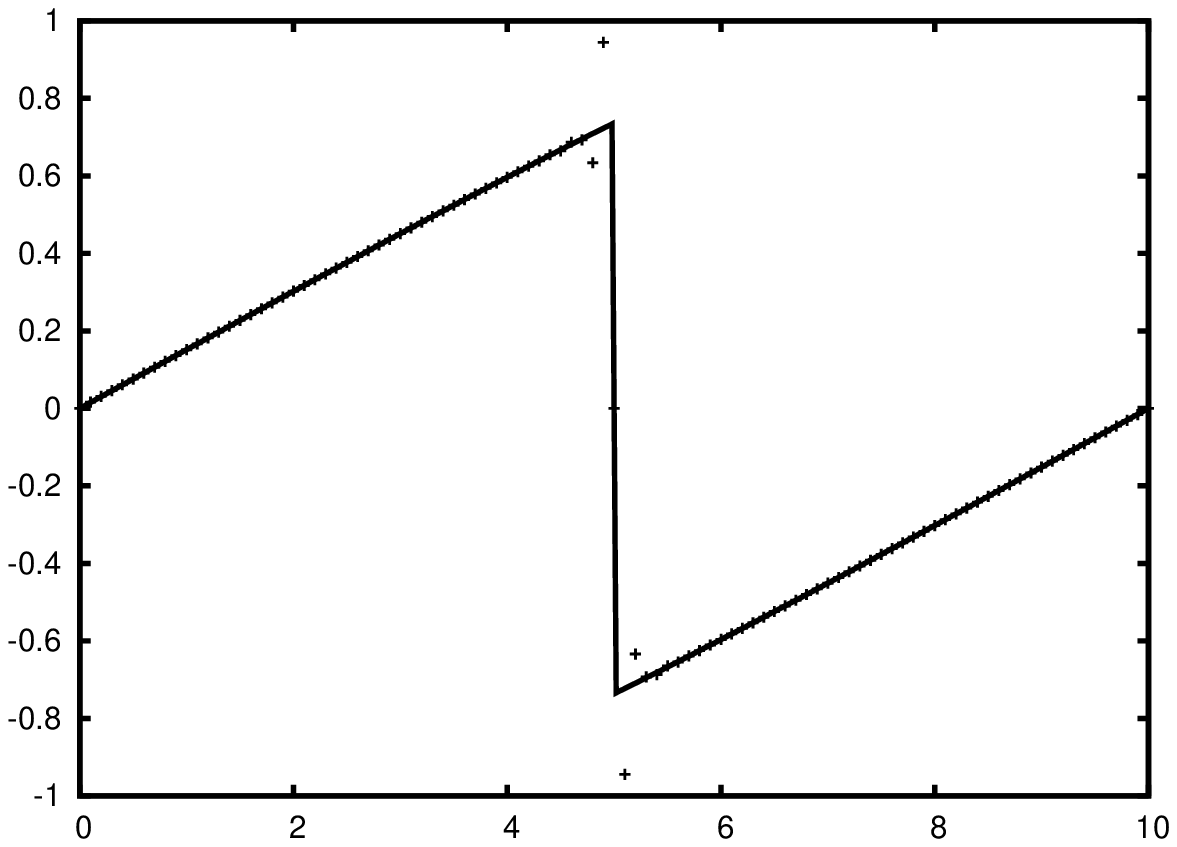}
\includegraphics[width=0.49\textwidth]{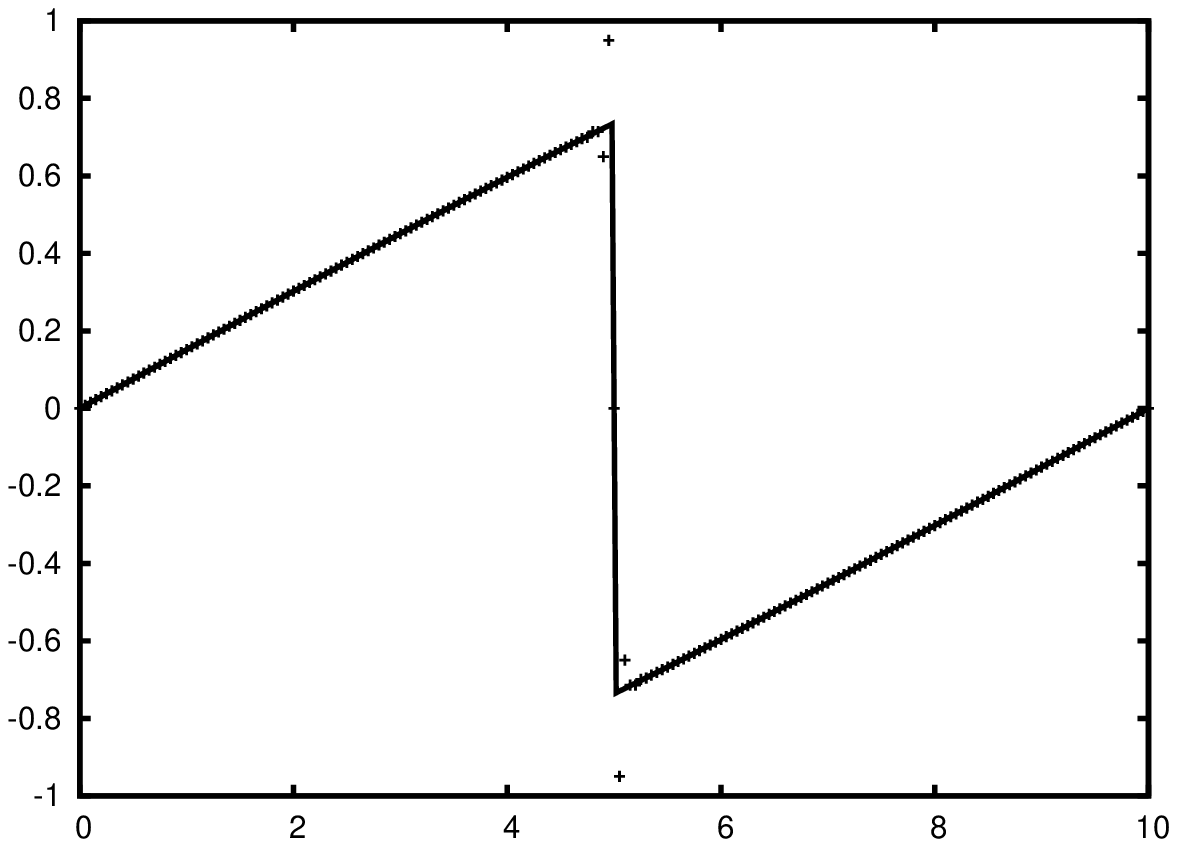}
\caption{Burgers equation: evolution of an initial sinus profile.
The numerical solution (point values) is plotted versus the exact
solution (continuous line), for $100$ points and $200$ points
resolution (left and right panels, respectively) and for the
($m=2$, $\beta=1/12$) and the ($m=3$, $\beta=2/75$) schemes (upper
and lower panels, respectively).} \label{burg}
\end{figure}

Burgers equation provides a simple example of a genuinely
non-linear scalar equation. A true shock develops from smooth
initial data. We will compute here the evolution of an initial
sinus profile, with fixed boundary conditions. We plot in
Fig.~\ref{burg} the numerical solution values versus (the
principal branch of) the exact solution, at the time where the
shock has fully developed. We compare $100$ points with $200$
points resolution (left and right panels, respectively), and also
the 3rd-order and 5th-order schemes described previously (upper
and lower panels, respectively). Concerning the resolution effect,
we can see here again that the spurious oscillations affect mainly
the points directly connected with the shock, in a number
depending on the stencil size but independent of the resolution.

\begin{figure}[t]
\centering
\includegraphics[width=0.49\textwidth]{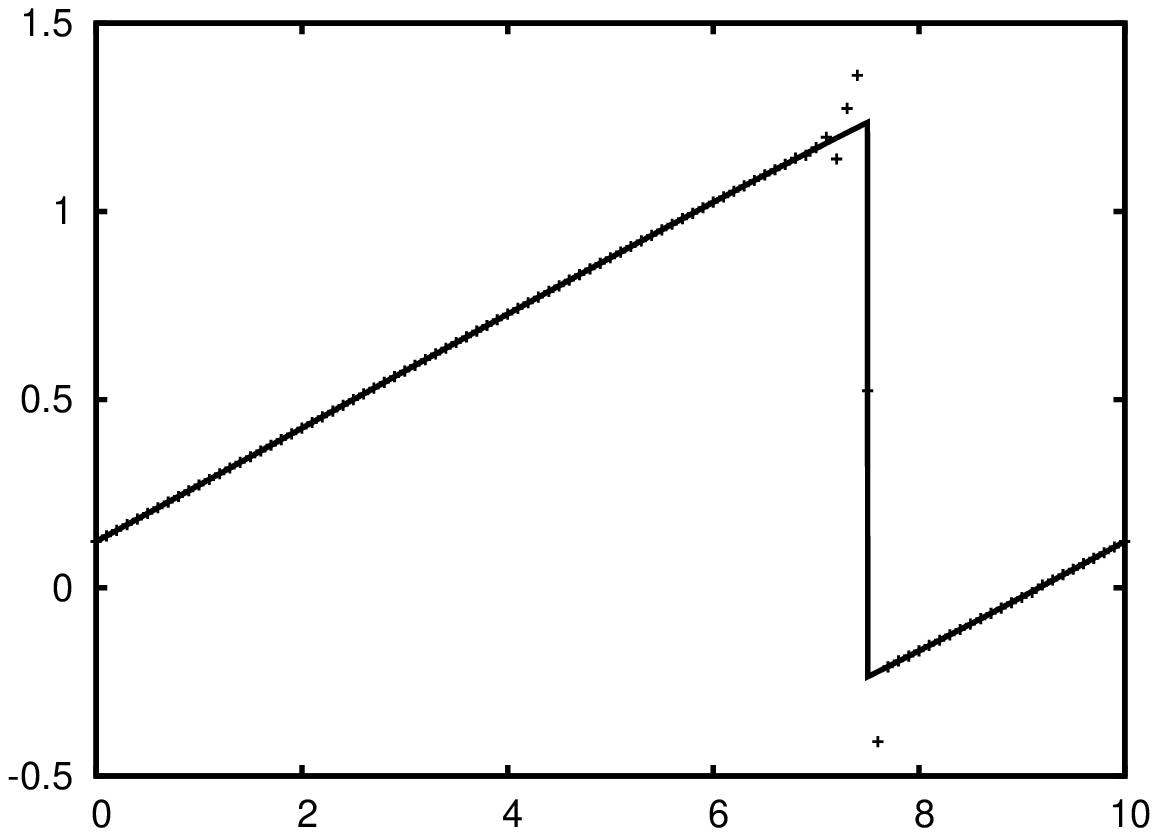}
\includegraphics[width=0.49\textwidth]{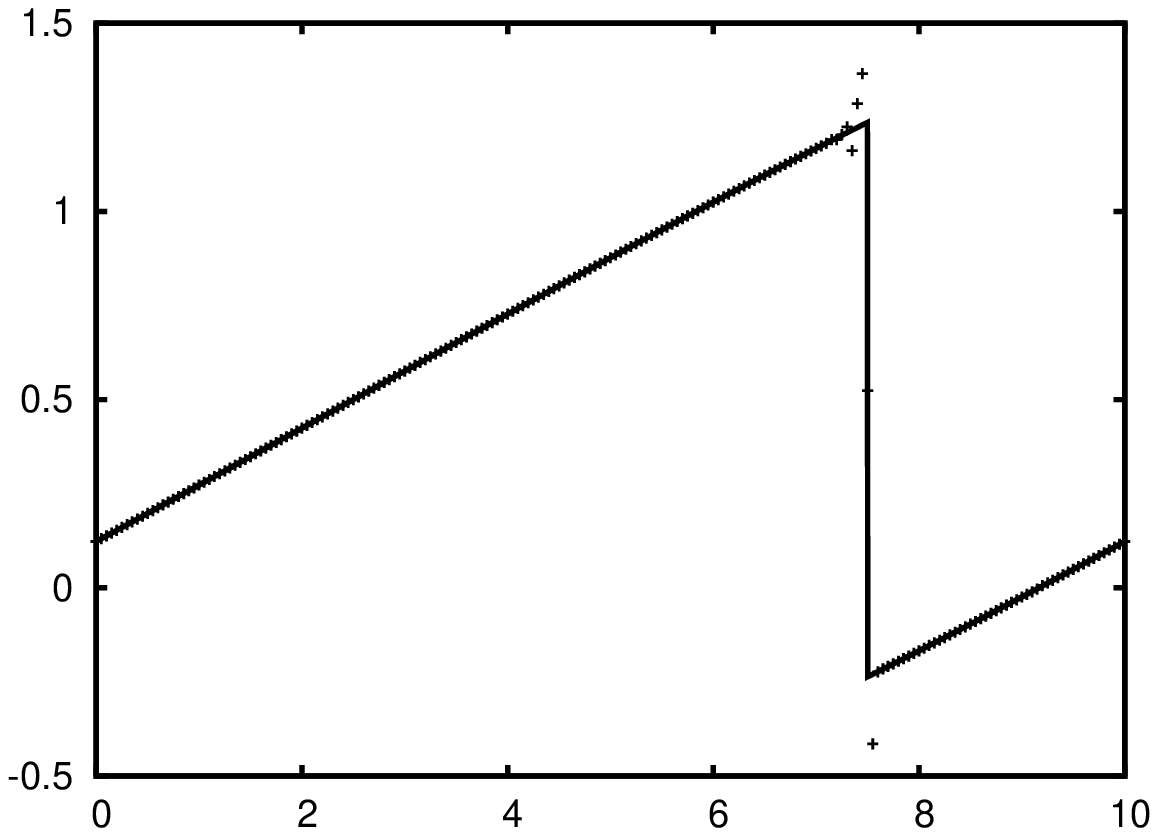}
\includegraphics[width=0.49\textwidth]{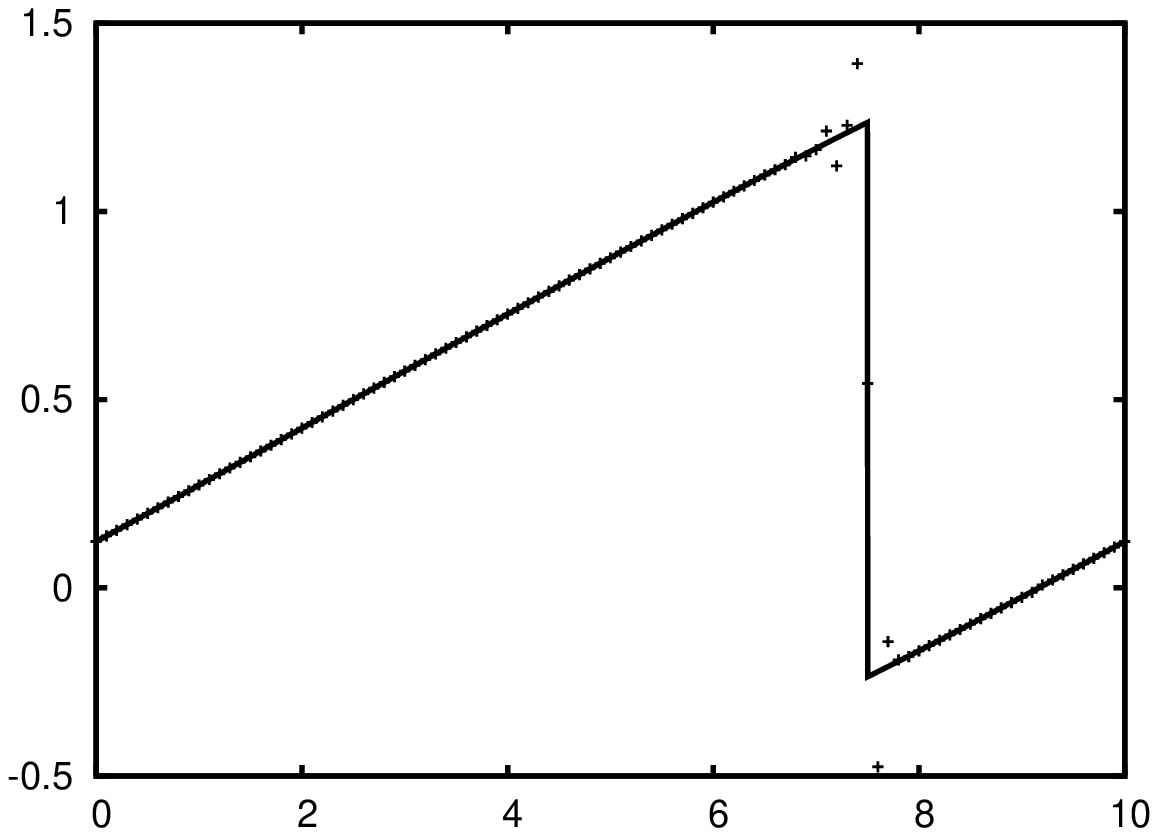}
\includegraphics[width=0.49\textwidth]{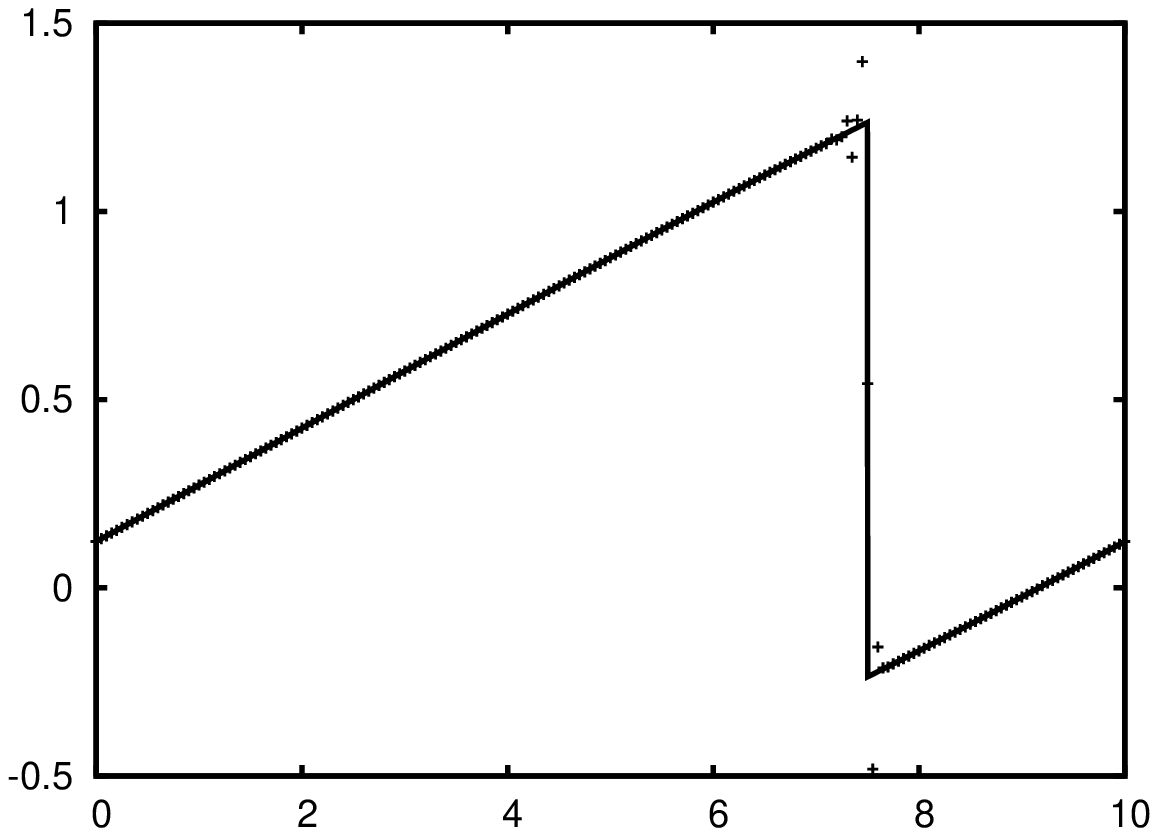}
\caption{Same as in the previous figure, but now for a moving
sinus profile. The numerical solution (point values) is plotted
versus the exact solution (continuous line), for $100$ points and
$200$ points resolution (left and right panels, respectively) and
for the ($m=2$, $\beta=1/12$) and the ($m=3$, $\beta=2/75$)
schemes (upper and lower panels, respectively).} \label{burg2}
\end{figure}

These conclusions are fully confirmed by a second simulation,
obtained by adding a constant term to the previous initial
profile, that is
\begin{equation}\label{burgers2}
    u(x) = \frac{1}{2} + sin(\frac{x\,\pi\,}{5})\,,
\end{equation}
with periodic boundary conditions. We can see in Fig.~\ref{burg2}
that a shock again develops, but it does no longer stand fixed: it
propagates to the right. Note that the plot shown corresponds to
$t=7$. We can confirm in this case that both the number of
spurious riddles and the magnitude of the overshots do not
increase with resolution, although it is larger in this case than
in the static shock one. We can confirm also that these effects
increase with the order-of-accuracy of the scheme: the larger
stencil adds one more riddle at every side and slightly larger
overshots.

These results clearly indicate convergence in the $L_1$ or similar
norms (but of course not in the maximum norm). Let us actually
perform a convergence test by considering the initial
profile~\cite{LO96}
\begin{equation}\label{burgers_conv}
    u(x, 0) = 1 + \frac{1}{2}\,sin(\pi\,x)\,,
\end{equation}
which is smooth up to $t=2/\pi$. We show in Table~\ref{table} the
errors at time $t=0.3$, where the shock has not yet appeared. The
first group of values corresponds to the third-order method, and
this is confirmed by the data both in the $L_1$ and the $L_\infty$
norms. The second group of values corresponds to the fifth-order
method, but only third-order accuracy is obtained from the
numerical values. This is because we keep using the third-order
Runge-Kutta algorithm (\ref{RK3}) for the time evolution. In order
to properly check the space discretization accuracy, we include a
third group of values, obtained with the same algorithm, but with
a much smaller time step in order to lower the time discretization
error: the leading error term is then due to the space
discretization and the expected fifth order accuracy is confirmed
by the numerical results, although the $L_\infty$ norm shows a
slightly decreasing convergence rate for the higher resolution
results.

\begin{table}[tbh]
\centering
\begin{tabular}{|c|c|c|c|c|}
\hline
Nx & $L_1$ error & $L_1$ order & $L_\infty$ error & $L_\infty$ order\\
\hline
160 & 7.22579 E-6 & 2.998 & 5.17334 E-5 & 2.981\\
\hline
320 & 9.04719 E-7 & 2.999 & 6.55306 E-6 & 2.994\\
\hline
640 & 1.13182 E-7 & 3.000 & 8.22735 E-7 & 2.998\\
\hline
1280 & 1.41486 E-8 &  & 1.03006 E-7 &  \\
\hline \hline
160 & 1.44981 E-6 & 3.017 & 9.57814 E-6 & 2.981\\
\hline
320 & 1.79043 E-7 & 3.005 & 1.21318 E-6 & 2.997\\
\hline
640 & 2.23035 E-8 & 3.003 & 1.51957 E-7 & 2.999\\
\hline
1280 & 2.78216 E-9 &    & 1.90041 E-8 &  \\
\hline\hline
160 & 7.09726 E-8 & 4.88 & 8.6567 E-7 & 4.97\\
\hline
320 & 2.41410 E-9 & 4.76 & 2.76804 E-8 & 3.98\\
\hline
640 & 8.92936 E-11 & 4.91 & 1.75192 E-9 & 3.48\\
\hline
1280 & 2.95859 E-12 &   & 1.36890 E-11 &  \\
\hline
\end{tabular}

\caption{Burgers problem. Norm of the errors and convergence rate
at $t=0.3$ for the initial profile (\ref{burgers_conv}). The first
group of values corresponds to the $m=2$ method with $\Delta
t=0.6\Delta x$. The second group corresponds to the $m=3$ method
with the same time step. The third group corresponds again to the
$m=3$ method, but with $\Delta t=0.06\Delta x$.} \label{table}
\end{table}

\subsection{Buckley-Leverett problem}

A more demanding test, still for the scalar case, is provided by
the Buckley-Leverett equation which models two-phase flows that
arise in oil-recovery problems~\cite{Leveque}. This equation
contains a non-convex (s-shaped) flux of the form
\begin{equation}\label{BLflux}
    f(u) = \frac{4 u^2}{4u^2+(1-u)^2}\,.
\end{equation}
Non-convex fluxes can lead to compound shock waves which are
shocks adjacent to a rarefaction wave with wave speed equal to the
shock speed at the point of attachment.

\begin{figure}[t]
\centering
\includegraphics[width=12cm, height=5cm]{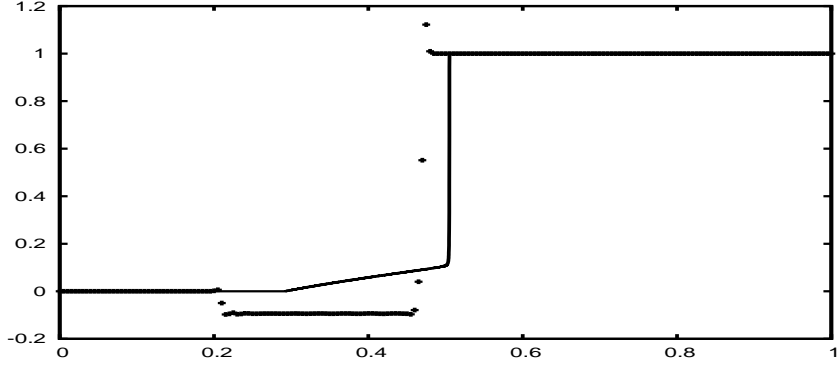}
\caption{Buckley-Leverett's problem. The continuous line
corresponds to the LLF first-order algorithm, with 10.000 points,
as a replacement for the exact solution. The crosses line
corresponds to the third-order algorithm ($m=2$, $\beta=1/12$)
with 200 points, converging towards a different solution.}
\label{BLplot}
\end{figure}

We will perform first a simulation with the initial data
\begin{equation}\label{BLdata}
u(x) = \left\{
 \begin{array}{cc}
  ~0 & \qquad 0 \leq \,x\, < \,1-1/\sqrt{2} \\
  ~1 & \qquad 1-1/\sqrt{2}\, \leq \,x\, < 1 \\
\end{array}
\right.
\end{equation}
so that the inflexion point in the flux (\ref{BLflux}) lies inside
the interval spanned by the data.

\begin{figure}[b]
\centering
\includegraphics[width=0.49\textwidth]{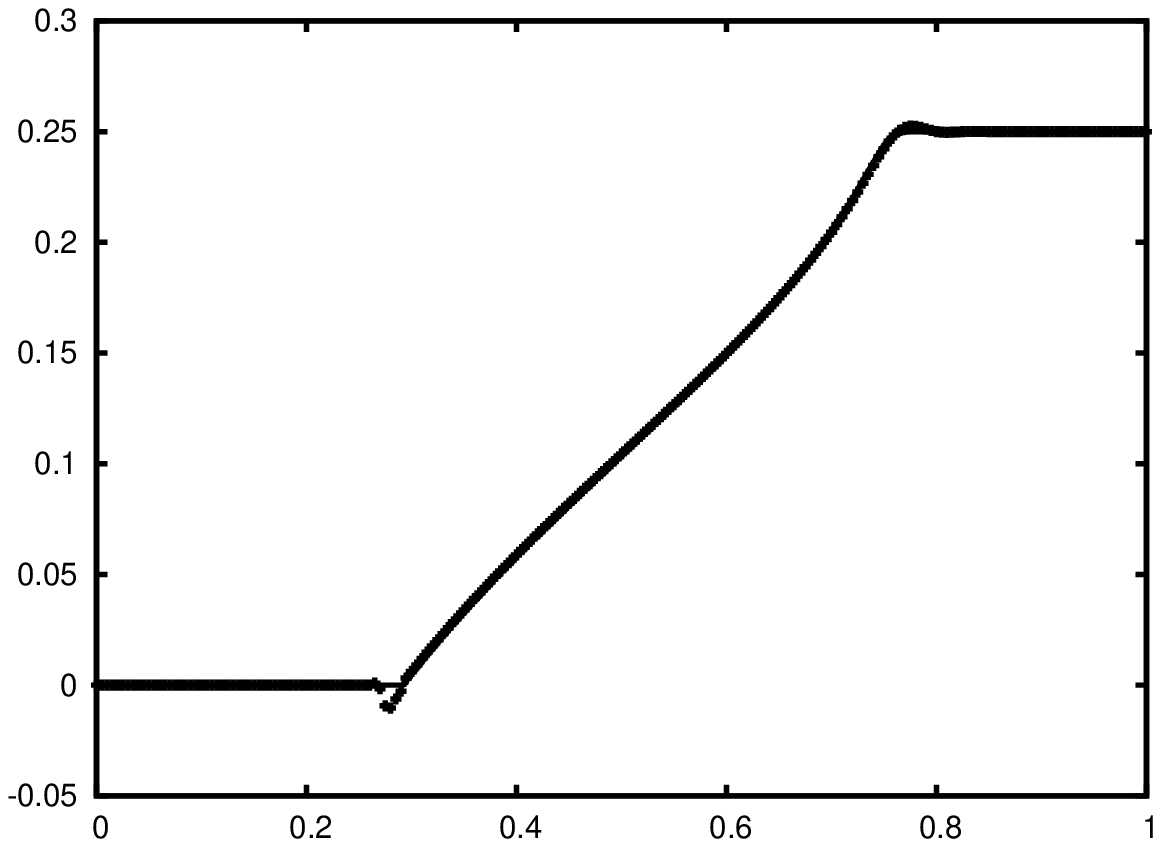}
\includegraphics[width=0.49\textwidth]{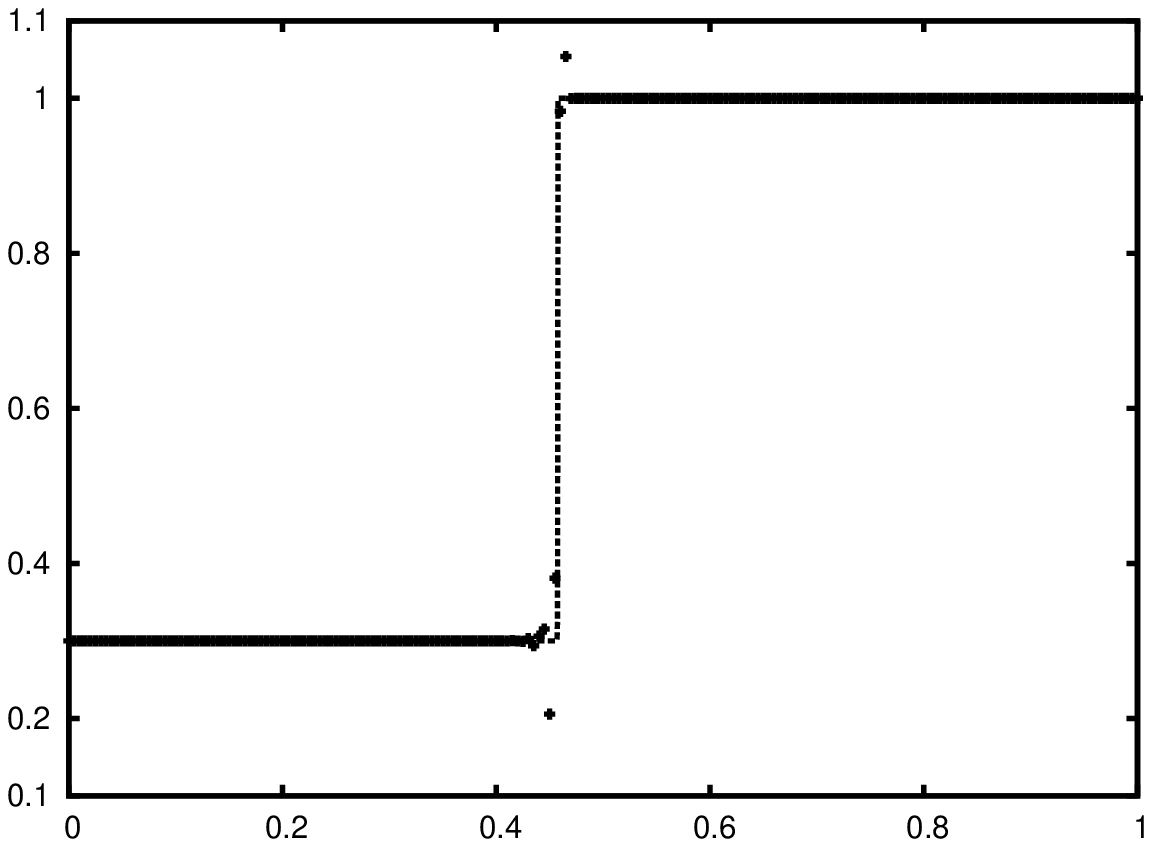}
\caption{Same as in the previous figure, but now for two different
dynamical ranges, which avoid the flux inflexion point. In the
left panel, an ordinary rarefaction wave appears, which is
correctly modelled by the third-order algorithm. In the right
panel, a simple shock appears, well captured by the third-order
algorithm. } \label{BLplot2}
\end{figure}

The exact solution in this case is well approximated by a
very-high-resolution (10.000 points) simulation using the
first-order LLF algorithm, as displayed in Fig.~\ref{BLplot}
(continuous line). We see a right-propagating compound shock wave,
consisting of a shock followed by a rarefaction wave, which
propagates in the same direction. The results for our third-order
algorithm, represented by the crosses line in Fig.~\ref{BLplot},
fail to reproduce correctly the rarefaction wave, which is
replaced by an spurious intermediate state, resulting into a
slower shock propagation speed.

In order to single out the problem, we have performed simulations
for the same flux (\ref{BLflux}) but with a dynamical range that
avoids the inflexion point either from below or from above. The
results are plotted in Fig.~\ref{BLplot2}, where we see either an
ordinary rarefaction wave (left panel) or a simple shock (right
panel), but no compound shock. In both cases, the third-order
algorithm ($m=2$, $\beta=12$) is able to model correctly the
dynamics. This results indicate that the problem with compound
shocks can be triggered by the presence of overshots at the
connection point between the shock and the associated rarefaction
wave, which can break the compound structure. The TVD character of
the LLF flux prevents this problem to arise, as it is clearly
shown in Fig.~\ref{BLplot} (continuous line).

\subsection{Euler equations}

\begin{figure}[b]
\centering
\includegraphics[width=0.49\textwidth]{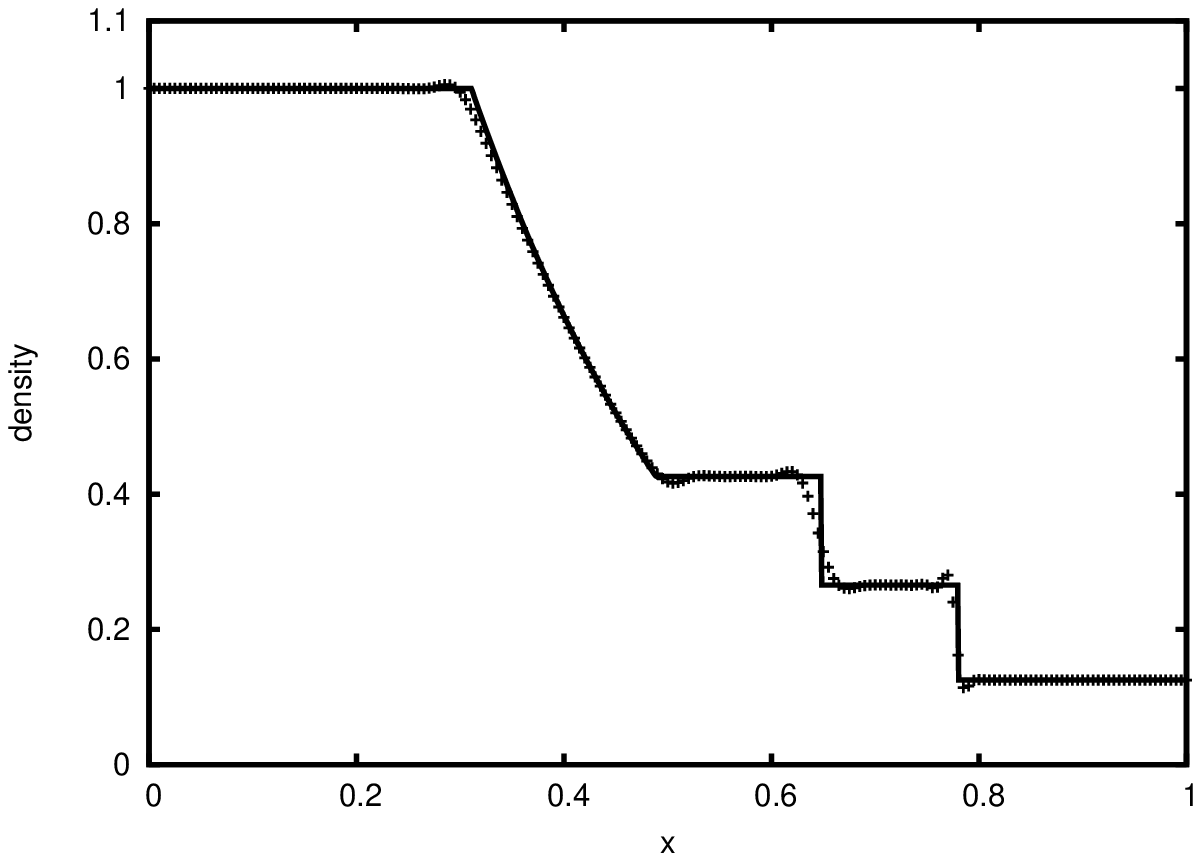}
\includegraphics[width=0.49\textwidth]{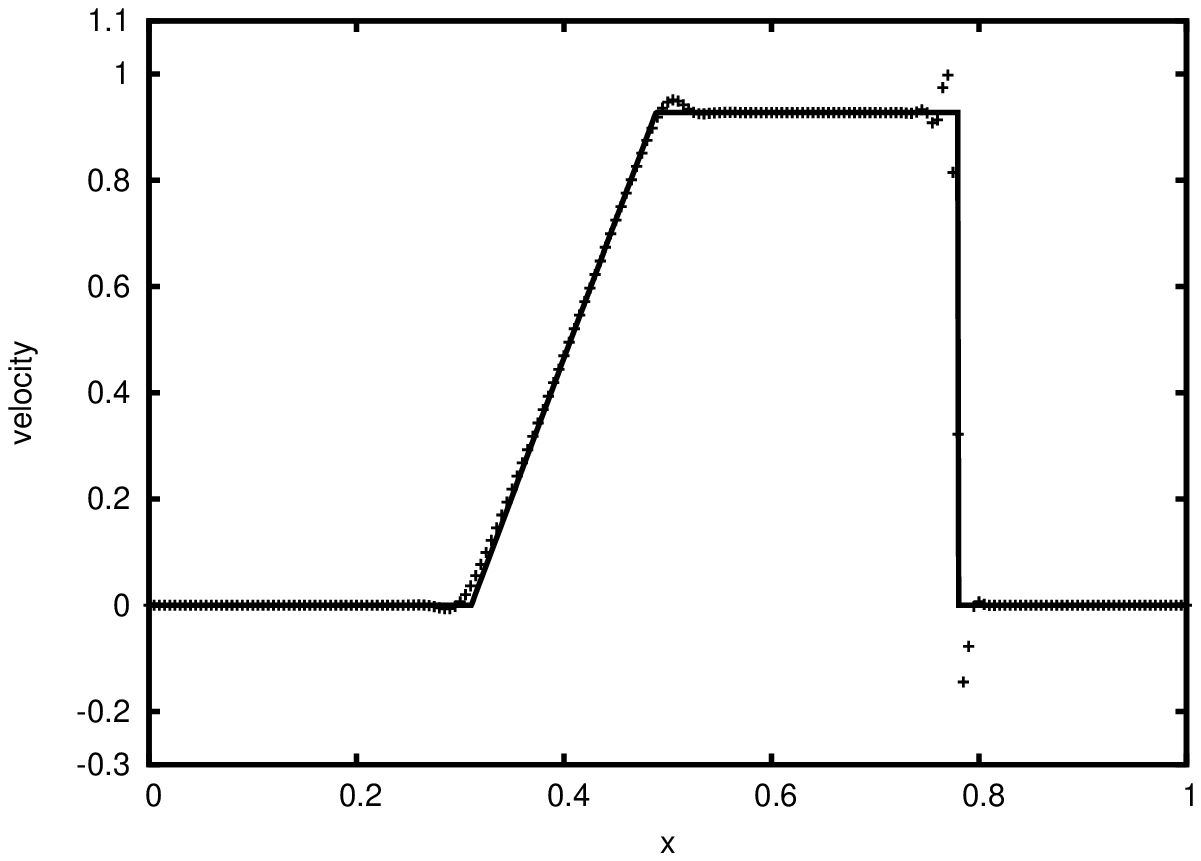}
\includegraphics[width=0.49\textwidth]{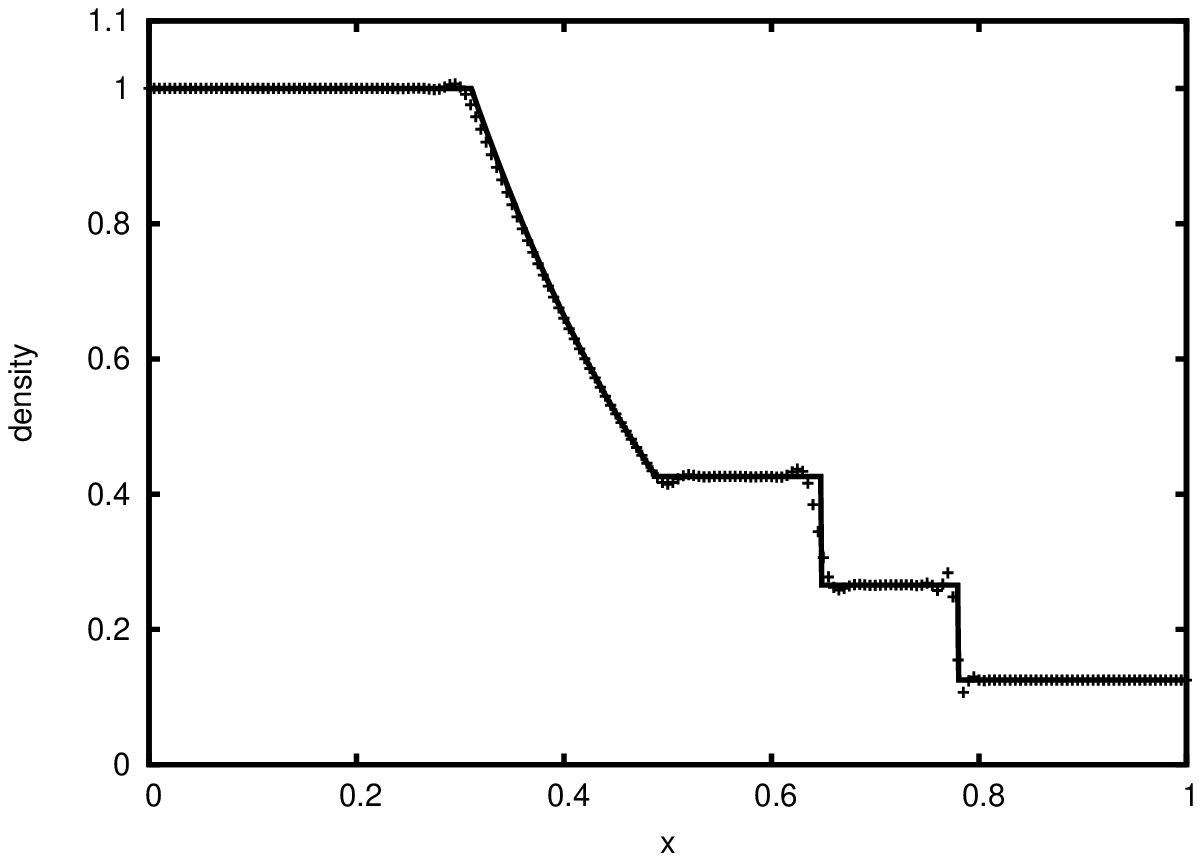}
\includegraphics[width=0.49\textwidth]{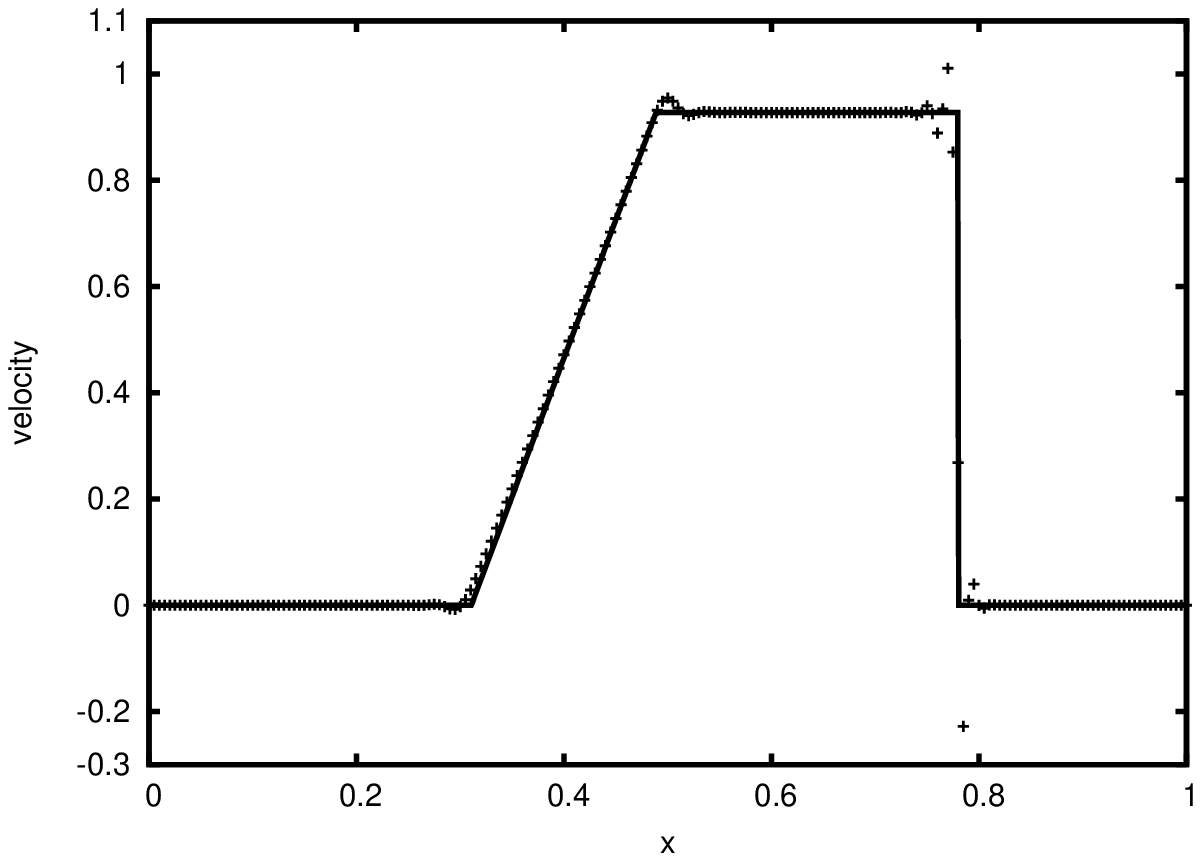}
\caption{Sod shock tube problem. Density and speed profiles (left
and right panels, respectively), for the ($m=2$, $\beta=1/12$) and
the ($m=3$, $\beta=2/75$) schemes (upper and lower panels,
respectively).} \label{sod}
\end{figure}

Euler equations for fluid dynamics are a convenient arena for
testing the proposed schemes beyond the scalar case. In the ideal
gas case, we can check the numerical results against well-known
exact solutions containing shocks, contact discontinuities and
rarefaction waves. We will deal first with the classical Sod
shock-tube test~\cite{Sod} with a standard $200$ points
resolution.

We plot in Fig.~\ref{sod} the gas density and speed profiles (left
and right panels, respectively). Looking at the 3rd-order scheme
results (upper panels), we see that both the rarefaction wave and
the shock are perfectly resolved, whereas the contact
discontinuity is smeared out. As a consequence, the main overshots
are just besides the shock, specially visible in the speed
profile, where the jump is much higher. Concerning the 5th-order
scheme (lower panels), the contact discontinuity is slightly
better resolved. This is however at the price of extra riddles and
more visible overshots, so that the 3rd-order scheme seems to be
more convenient.

\begin{figure}[h]
\centering
\includegraphics[width=0.49\textwidth]{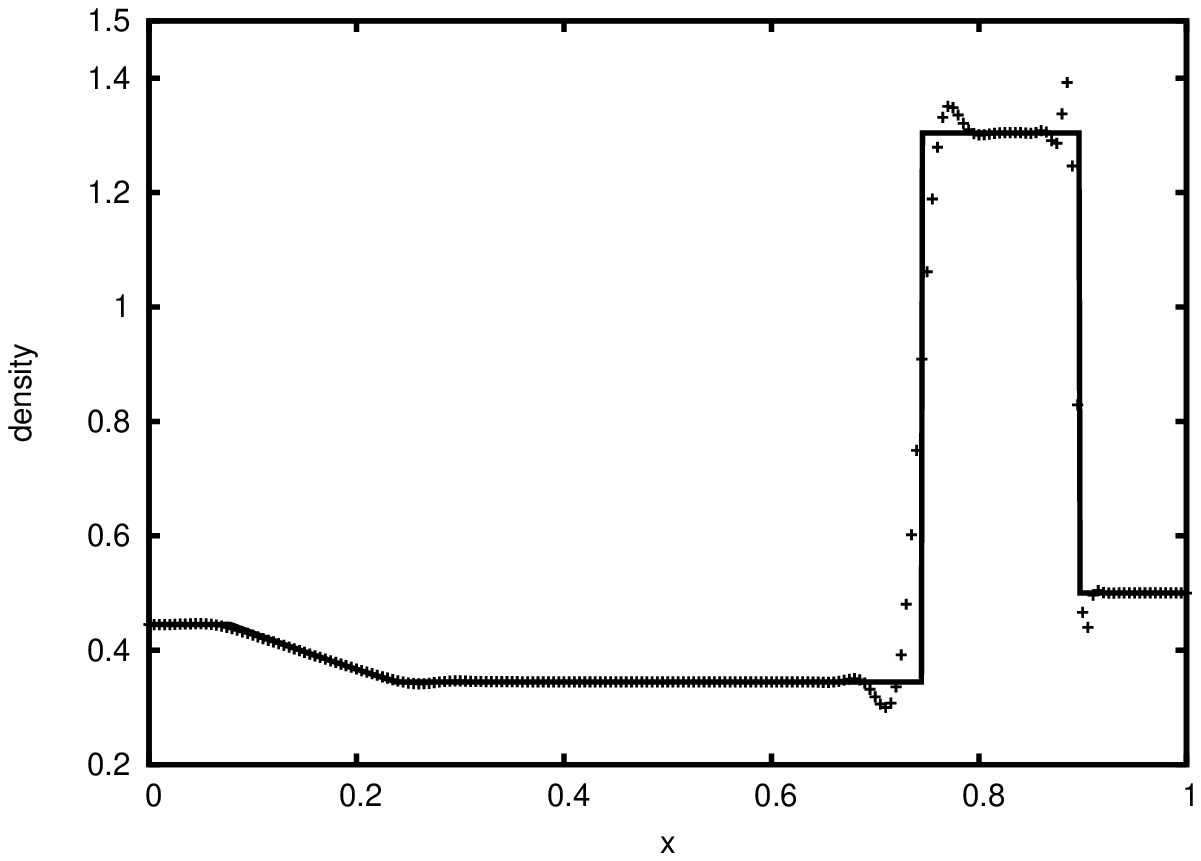}
\includegraphics[width=0.49\textwidth]{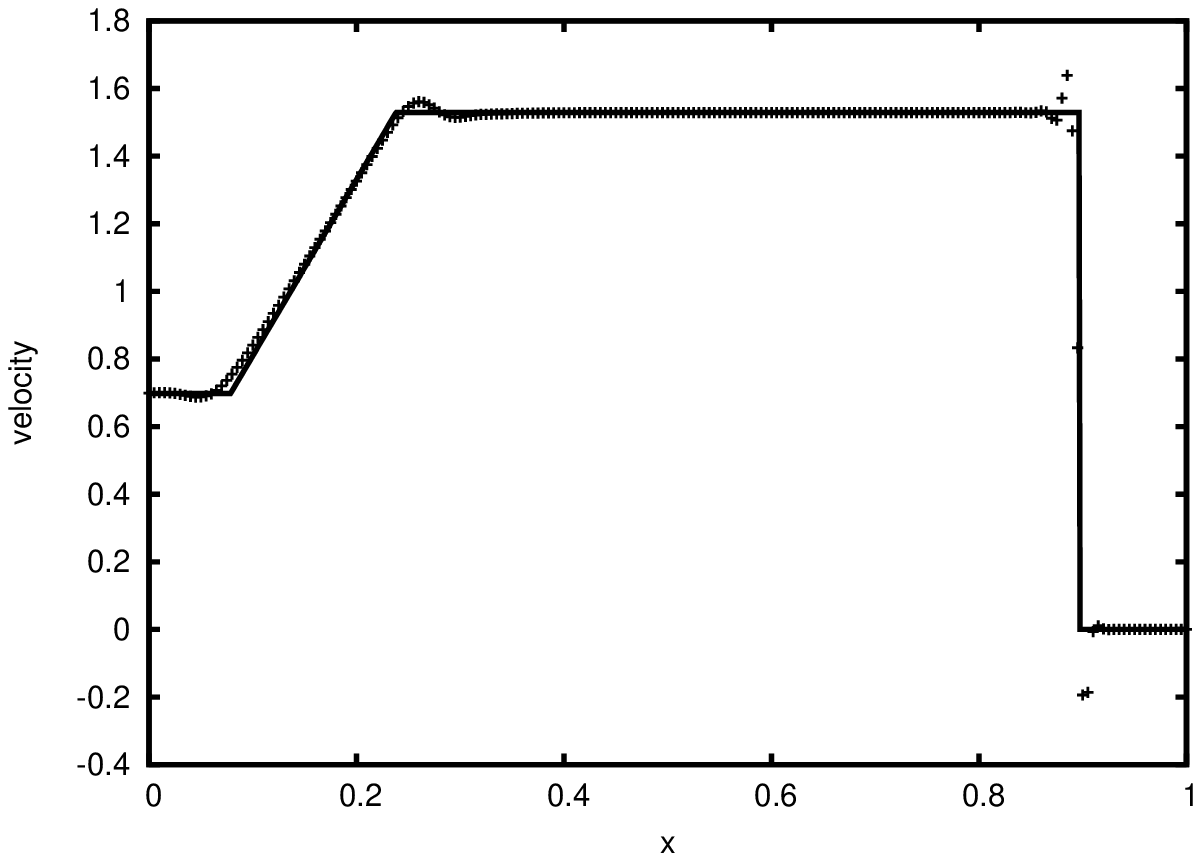}
\includegraphics[width=0.49\textwidth]{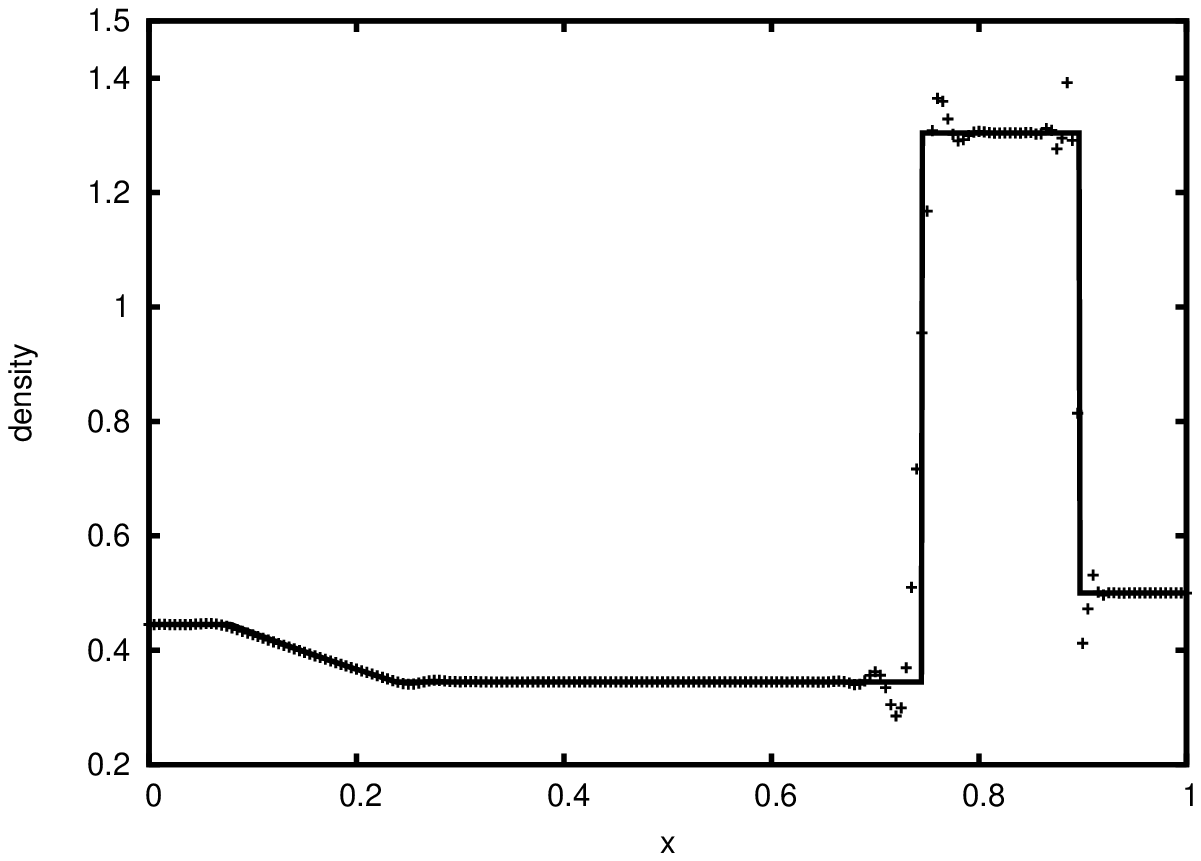}
\includegraphics[width=0.49\textwidth]{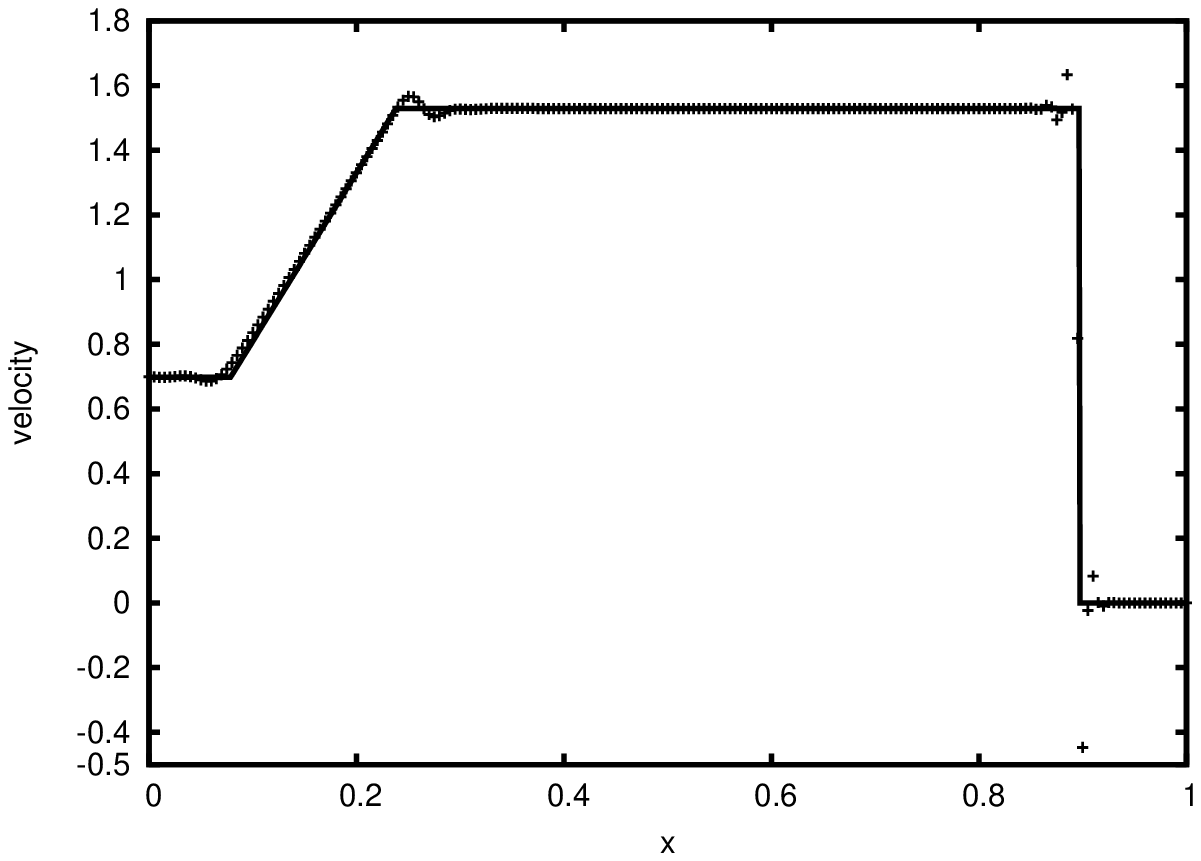}
\caption{Lax shock tube problem. Density and speed profiles (left
and right panels, respectively), for the ($m=2$, $\beta=1/12$) and
the ($m=3$, $\beta=2/75$) schemes (upper and lower panels,
respectively).} \label{lax}
\end{figure}

A more demanding test is obtained when assuming a discontinuity in
the initial speed, as in the Lax test~\cite{Lax_test}. As we see
in Fig.~\ref{lax}, we get the same behavior than for the Sod test
case. The main difference is that the density jump at the contact
discontinuity is much higher: the smearing of the density profile
there is more visible, in contrast with the sharp shock profile
nearby. Note also that some speed overshots are greater than the
ones arising in the Sod test case (we have kept here the same
$200$ points resolution for comparison).  The third-order
algorithm seems to be more convenient again in this case.

\section{Multidimensional tests}
\label{2-3D}The results of this paper can be extended to a
multidimensional case in a simple way. The semi-discrete equation
(\ref{Flux_deriv}) can be written in a rectangular grid as
follows:
\begin{equation}\label{Flux_deriv2D}
    \partial_t u_{i, j} = - \frac{1}{\Delta x}~(f_{i+1/2, j}-f_{i-1/2, j})
    ~- \frac{1}{\Delta y}~(f_{i, j+1/2}-f_{i, j-1/2}),
\end{equation}
and the numerical flux can be computed by applying
(\ref{flux_develop}) to every single direction. Note however that
the restriction (\ref{newcourbound}) on the time step must be
extended in this case to
\begin{equation}\label{newcourbound2}\hspace{-0.3cm}
       \lambda_j~\Delta t~(\frac{1}{\Delta x}+\frac{1}{\Delta y})
       ~[~d^{~m}_{-1}-d^{~m}_{0}+b~\sum_{k\neq 0}
    max(d^{~m}_{k-1}-d^{~m}_{k},0)~] ~\le~
    1/2\,.
\end{equation}

In the finite-difference version (\ref{Dform_new}), the extension
to the multidimensional case amounts to replicate the
right-hand-side difference operators for every single direction:
no cross-derivative terms are required. This multidimensional
extension allows to deal with some MHD tests, which add more
complexity to the dynamics, clearly beyond the simple tests
considered in the previous section.

\subsection{The Orszag-Tang 2D vortex problem}

As a first multi-dimensional example, let us consider here the
Orszag-Tang vortex problem~\cite{OT}. This is a well-known model
problem for testing the transition to supersonic
magnetohydrodynamical (MHD) turbulence and has become a common
test of numerical MHD codes in two dimensions.

\begin{figure}[h]
\centering
\includegraphics[width=0.49\textwidth]{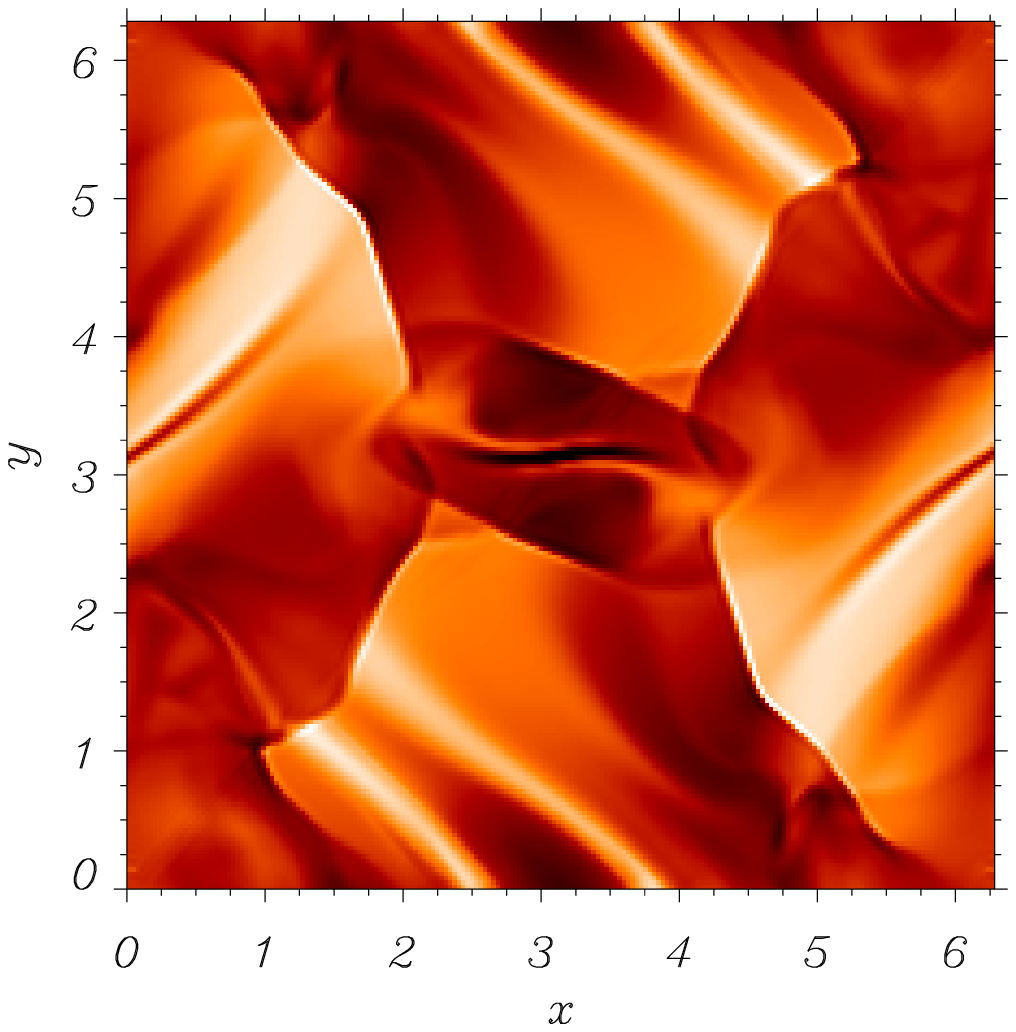}
\includegraphics[width=0.49\textwidth]{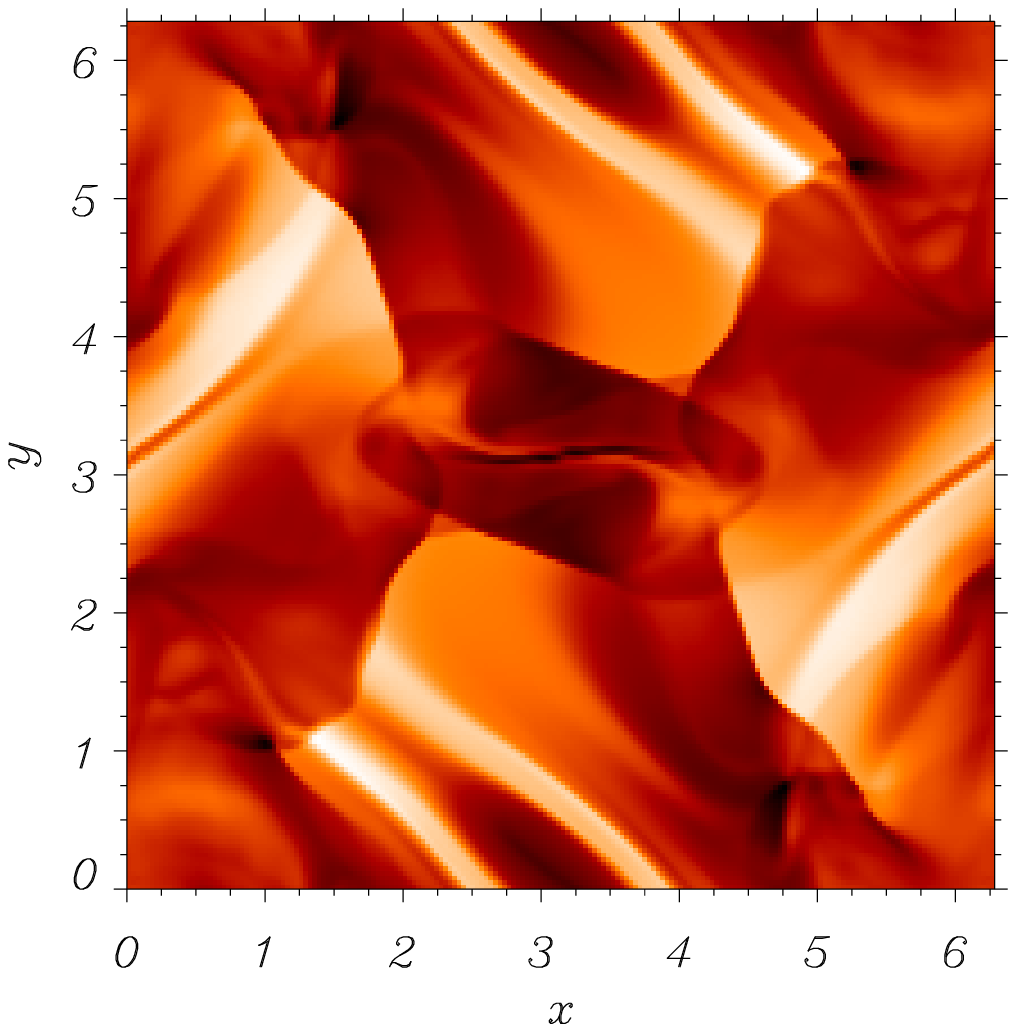}
\caption{Temperature at $t=3.14$ in the Orszag-Tang vortex test
problem. In this simulation the grid has $200\times 200$ mesh
points. In the left panel the third-order scheme ($m=2$,
$\beta=1/12$) has been used while in the right panel the result is
for a a second order scheme built from the Roe-type solver and the
MC limiter.} \label{ot}
\end{figure}

A barotropic fluid ($\gamma=5/3$) is considered in a doubly
periodic domain $\left[0,2\pi\right]^2$, with uniform density
$\rho$ and pressure $p$. A velocity vortex given by ${\bf
v}=(-\sin y, \sin x)$, corresponding to a Mach 1 rotation cell, is
superimposed with a magnetic field ${\bf B}=(-\sin y, \sin 2x)$,
describing magnetic islands with half the horizontal wavelength of
the velocity roll. As a result, the magnetic field and the flow
velocity differ in their modal structures along one spatial
direction.

In Fig.~\ref{ot} (left panel) the temperature, $T=p/\rho$, is
represented at a given time instant ($t=3.14$). The figure clearly
shows how the dynamics is an intricate interplay of shock
formation and collision. The numerical scheme, with $m=2$ and
$\beta=1/12$ seems to handle the Orszag-Tang problem quite well.
In Fig.~\ref{ot} (right panel) we plot the results for the same
problem using a second order scheme built from the Roe-type solver
and the monotonized-central (MC) symmetric limiter~\cite{MC}. The
results with both methods are qualitatively very similar.

\subsection{Torrilhon MHD shock tube problem}

We now consider the MHD shock tube problem described by
Torrilhon~\cite{Torr} to investigate dynamical situations close to
critical solutions. We will assume again a barotropic fluid with
$\gamma=5/3$. The initial conditions for the components of the
magnetic field $(B_2,B_3)$ are $(\cos\theta,\sin \theta)$, with
$\theta=0$ for $x\leq 0$. Depending on the angle $\theta$ between
the left and right transverse components of the magnetic field,
different types of solutions are found. Regular $r$-solutions
consist only of shocks or contact discontinuities. Critical
$c$-solutions appear in the coplanar case, where the angle
$\theta$ is an integer multiple of $\pi$. These solutions can
contain also non-regular waves, such as compound waves.

\begin{figure}[t]
\centering
\includegraphics[width=10cm]{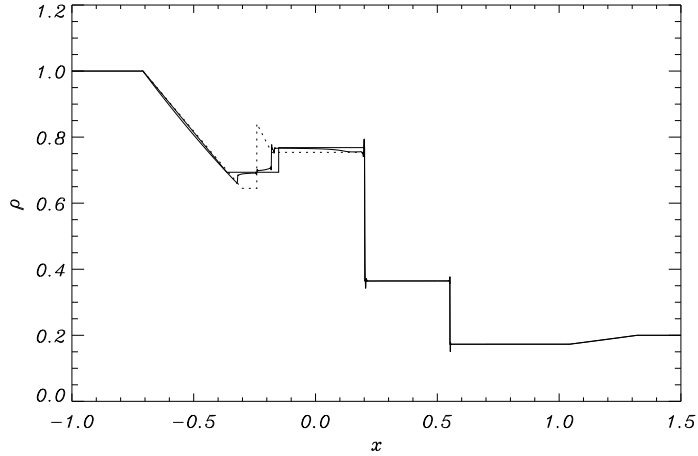}
\caption{Plot of the density $\rho$ at $t=0.4$ for the almost
co-planar problem with $\theta=3$. In this simulation $5000$ mesh
points have been used. The dashed line represents the critical
$c$-solution while the solid black line is the correct
$r$-solution. Both solutions differ clearly in the interval
$[-0.35,-0.1]$. The numerical simulation lies between the two.}
\label{torri}
\end{figure}

We consider the situation for an $almost$ co-planar case,
$\theta=3$. Analytically, this has a regular $r$-solution, but the
numerical solution is attracted towards the nearby critical
solution for $\theta=\pi$. Fig.~\ref{torri} shows the density
profile plotted together with the correct $r$-solution (solid
black line) and the co-planar $c$-solution (dashed line). The
$r$-solution has, from left to right, a rarefaction, a rotation, a
shock, a contact discontinuity, a shock, a rotation and a
rarefaction. The discrepancies among the different solutions are
mainly in the interval $[-0.35,-0.1]$.

\begin{figure}[t]
\centering
\includegraphics[width=10cm]{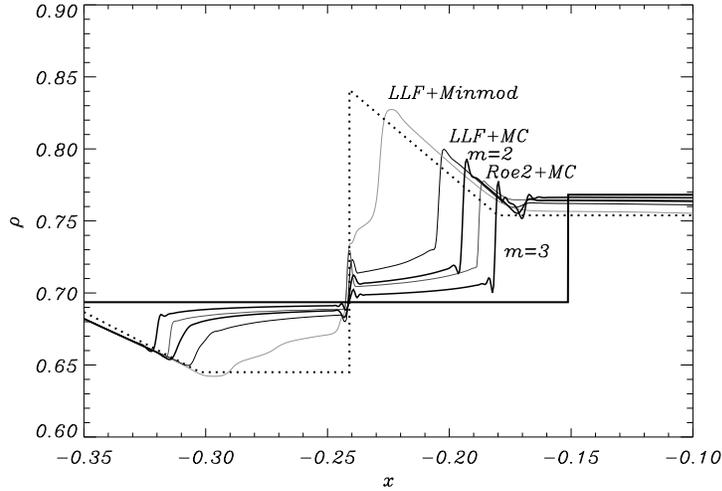}
\caption{Same as Fig.~\ref{torri}, but enlarging the interval
where the discrepancies show up. In addition to the exact regular
and critical solutions, we plot, from top to bottom, the
simulations for schemes using LLF with minmod limiter, LLF with MC
limiter, the unlimited $m=2$ algorithm, a Roe solver with MC
limiter and the unlimited $m=3$ algorithm.} \label{torriclose}
\end{figure}

This interval is magnified in Fig.~\ref{torriclose}. The solid
black line is the correct $r$-solution while the dashed line
represents the critical $c$-solution. We see that the solutions
with $m=2$ and $m=3$ tend to the correct solution although they
keep some remnant from the $c$-solution. For comparison purposes
we have also represented the numerical solution obtained with
other schemes. We have used a second order LLF scheme and a second
order Roe solver with either the minmod or the MC slope limiters.

The LLF scheme with the minmod limiter gets too close to the
$c$-solution, even for this high-resolution simulation. The
situation improves by replacing the minmod limiter by the MC one,
but still gets farther from the right solution than the schemes
proposed in this paper. Only the combination of a Roe-type solver
with the MC limiter improves the results of the third-order scheme
($m=2$), but not those of the fifth-order scheme ($m=3$). This
problem provides one specific example in which the fifth-order
scheme seems to be more convenient than the third order one: the
extra riddles are actually compensated by a clear improvement in
the solution accuracy.

\subsection{Double Mach reflection problem}
This problem is a standard test case for high-resolution schemes.
It corresponds to an experimental setting in which a shock is
driven down a tube which contains a wedge. We will adopt here the
standard configuration proposed by Woodward and
Colella~\cite{WoodCol84}, involving a Mach 10 shock in air
($\gamma=1.4$) at a $60^o$ angle with a reflecting wall. The air
ahead of the shock is stationary with a density of 1.4 and a
pressure of 1. The reflecting wall lies at the bottom of the
computational domain, starting at $x=1/6$. Allowing for this, the
exact post-shock condition is imposed at the bottom boundary in
the region $0\leq x \leq 1/6$ and a reflecting wall condition is
imposed for the rest. Inflow (post-shock) conditions are used at
the left and top boundaries, whereas an outflow (no gradient)
condition is used for the right boundary.

\begin{figure}[h]
\centering
\includegraphics[width=15cm, height=7cm]{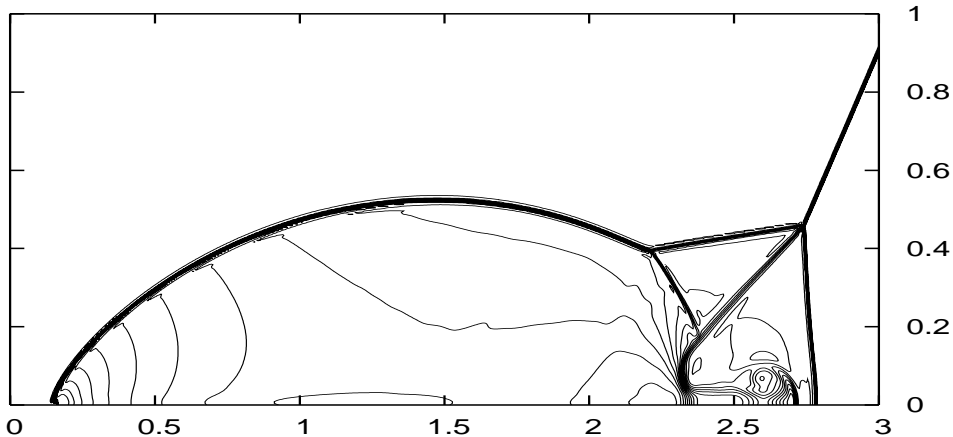}
\caption{Double Mach reflection. Density plot at $t=0.2$. The
simulation is made with the 3rd-order method ($m=2$, $\beta=1/12$)
with $\Delta x =\Delta y = 1/240$. 30 evenly spaced density
contours are shown.} \label{2Machbig}
\end{figure}

\begin{figure}[t]
\centering
\includegraphics[width=0.49\textwidth]{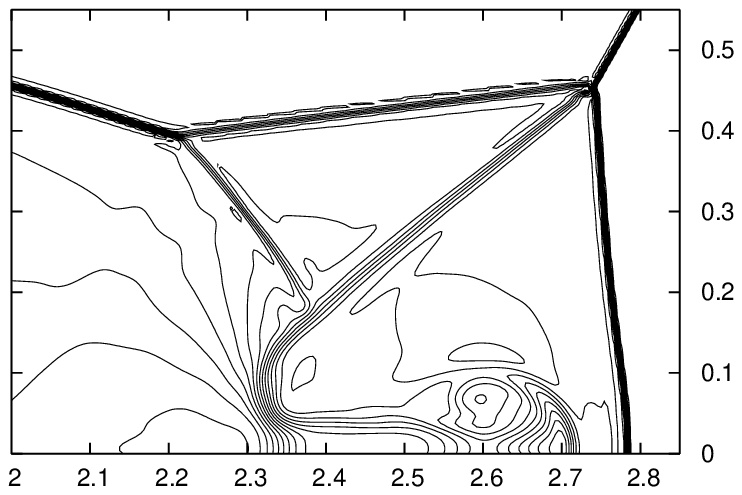}
\includegraphics[width=0.49\textwidth]{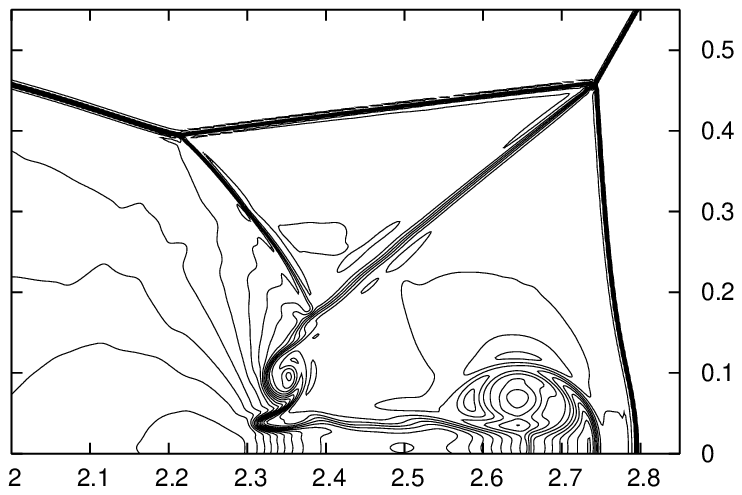}
\includegraphics[width=0.49\textwidth]{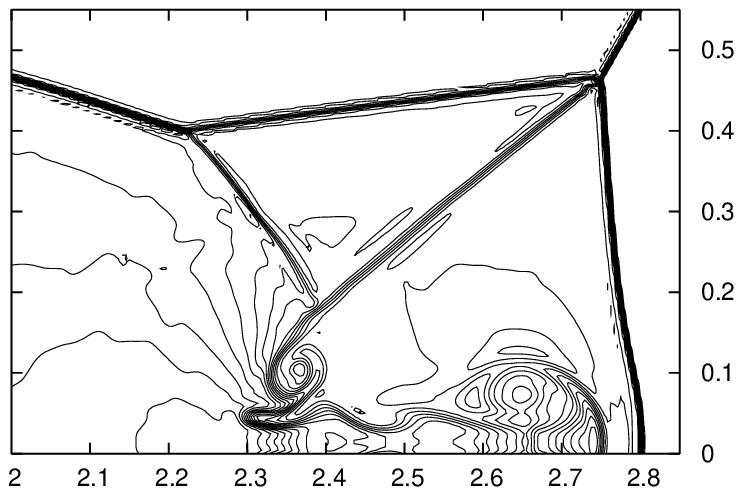}
\includegraphics[width=0.49\textwidth]{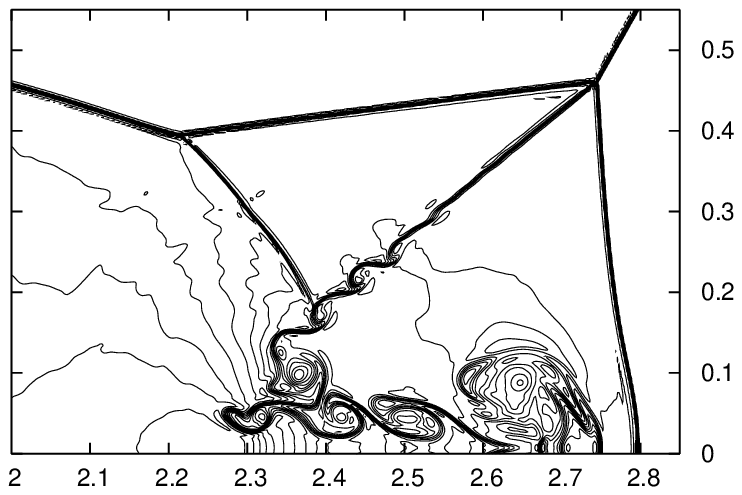}
\caption{Same as Fig.~\ref{2Machbig}, but enlarging the lower
right corner. The top panels correspond to the third-order method
($m=2$, $\beta=1/12$), with resolution of either $1/240$ (left) or
$1/480$ (right). The bottom panels show the same for the 5th-order
method ($m=3$, $\beta=2/75$). Both the jet near the bottom wall
and the weak shock, generated at the kink in the main reflected
shock, are well resolved. A vortex structure at the bottom of the
diagonal contact discontinuity shows up, with more details
appearing when increasing accuracy.} \label{2Machcloseup}
\end{figure}

This configuration leads to a complex flow structure, produced by
a double Mach reflection of the shock at the wall. A self-similar
flow develops as the fluid meets the reflecting wall. Two Mach
stems develop, with two contact discontinuities. We have plotted
in Fig.~\ref{2Machbig} the density contours at $t=0.2$, when the
main features have fully developed. The more challenging ones are
the jet propagating to the right near the reflecting wall and the
weak shock generated at the second Mach reflection, as seen in the
enlarged area in Fig.~\ref{2Machcloseup}.

Our third-order results agree with the original
ones~\cite{WoodCol84} for the corresponding resolution: both the
jet and the weak shock are clearly captured. Increasing both the
resolution and the order-of-accuracy of the numerical algorithm,
as shown in the subsequent panels in Fig.~\ref{2Machcloseup}, we
see more details of the jet rolling-up. Also, a vortex structure
appears near the bottom wall, which starts affecting the diagonal
contact discontinuity arising from the triple point. These
high-resolution features, appearing in the last panel in
Fig.~\ref{2Machcloseup}, agree with the ones obtained with a WENO
method of the same order (but double resolution, $1/960$) in
Ref.~\cite{SZS03}. This also agrees with the results of recent
spectral (finite) volume simulations~\cite{WZL04}, in which those
structures show up gradually, as one is getting more accurate
simulations. This is another example in which a higher-order
algorithm can be preferred, as it captures more detailed features
of complex structures for a given resolution.

\section{Conclusions and outlook}
The numerical experiments presented in this paper provide clear
evidence for a TVB behavior of the proposed schemes. This means
that the total variation growth is uniformly bounded,
independently of the resolution, for a fixed evolution time.
Moreover, the experimental pattern is a sudden growth of the total
variation, which provides a time-independent bound for the rest of
the evolution. This growth is confined to the mesh points directly
connected with non-sonic critical points, especially near
discontinuities. But the resulting riddles do not spread over
smooth regions and the overall features of the solution are
preserved as a result. In the case of compound shocks, however,
the numerical simulations actually mystify the physical solution:
the spurious riddles affect the contact point between the shock
and the adjacent rarefaction wave, breaking the compound
structure, even if the TVB behavior is still preserved.

The proposed schemes are obtained from the unlimited version of
the Osher-Chakrabarthy~\cite{ICASE} linear flux-modification
algorithms. The robustness of the unlimited version is related
with the high compression factor of this algorithms family. We
have actually improved the available estimates up to a remarkable
value of $b=5$, for the third-order case. This suggests that these
estimates could be even improved by using alternative
bound-setting procedures. Unfortunately, even in the scalar case,
we are not able to prove rigorously the TVB properties of these
methods, although we are currently working in this direction.

We have combined the unlimited Osher-Chakrabarthy algorithm with
the simple LLF flux formula. As a result, we have been able to
derive the compact finite-difference scheme (\ref{Dform_new}),
which is equivalent to the corresponding finite-volume
implementation in the unlimited case. This provides an extremely
cost-efficient algorithm for dealing with the most common
problems, even in presence of interacting dynamical shocks, as we
have done in the Orszag-Tang 2D vortex and the double Mach
reflection cases. Of course, its use should be limited to
convex-flux problems, where compound shocks do not arise.

The resulting finite-difference formula (\ref{Dform_new}) is
similar to the ones obtained by the 'artificial viscosity'
approach (see for instance ref.~\cite{KO}). The main difference is
that the spectral radius plays a key role here in the dissipation
term, providing some sort of 'adaptive viscosity'. Even in the
most simple constant-speed case we can still see that the order of
accuracy in our case is dictated by the dissipation term, in
contrast with the extra freedom one gets in the standard
artificial viscosity approach. Moreover, our compression factor
estimates provide specific prescriptions for the value of the
dissipation coefficient.




\begin{thebibliography}{00}

\bibitem{Lax_Wendroff}
P.~D.~Lax, B.~Wendroff (1960), "Systems of conservation laws".
Commun. Pure Appl Math. 13: 217-237.

\bibitem{MacCormack}
R.~W.~MacCormack (1969), "The Effect of viscosity in hypervelocity
impact cratering". AIAA Paper 69-354.

\bibitem{Godunov}
S.~K.~Godunov (1959), "A Difference Scheme for Numerical Solution
of Discontinuos Solution of Hydrodynamic Equations". Math. Sbornik
47: 271-306, translated US Joint Publ.~Res.~Service, JPRS 7226,
(1969).

\bibitem{Rusanov}
V.~V.~Rusanov (1961), "Calculation of Intersection of Non-Steady
Shock Waves with Obstacles". J.~Comput.~Math.~Phys. USSR 1:
267-279.

\bibitem{Harten83}
A.~Harten (1983), "High Resolution Schemes for Hyperbolic
Conservation Laws". J.~Comput.~Phys. 49: 357-393.

\bibitem{ICASE}
S.~Osher and S.~Chakravarthy (1984), "Very High Order Accurate TVD
schemes", ICASE Report 84-44, IMA Volumes in Mathematics and its
Applications vol 2: 229-274. Springer-Verlag, 1986.

\bibitem{Leveque}
R.~J.~LeVeque (1992), "Numerical Methods for Conservation Laws",
Lectures in Mathematics, Birkh\"{a}user.

\bibitem{Shu87}
Chi-Wang Shu (1987), "TVB Uniformly High-Order Schemes for
Conservation Laws", Mathematics of Computation 49:105-121.

\bibitem{ENOa}
A.~Harten and S.~Osher (1987), "Uniformly high-order accurate
nonoscillatory schemes, I", SIAM J.~Num.~Anal.~24:279-309

\bibitem{ENOb}
A.~Harten, B.~Engquist, S.~Osher and S.~Charavarty (1987),
Uniformly high-order accurate essentially non-oscillatory
schemes", J.~Comp.~Phys.~71:231-303.

\bibitem{Balsara-Shu}
D.~S.~Balsara and Chi-Wang Shu (2000), "Monotonicity preserving
weighted essentially non-oscillatory schemes with increasingly
high order of accuracy", J.~Comp.~Phys.~160:405-452.

\bibitem{LPR00}
Doron Levy, Gabriella Puppo and Giovanni Russo (2000), "A third
order central WENO scheme for 2D conservation laws", Applied
Numerical Mathematics 33:415.

\bibitem{BL06}
Steve Bryson and Doron Levy (2006), "On the total variation of
High-Order Semi-Discrete Central schemes for Conservation Laws",
Journal of Scientific Computation 27:163.

\bibitem{KT00}
A.~Kurganov and E.~Tadmor (2000), "New High-Resolution Central
Schemes for Nonlinear Conservation Laws and Convection-Diffusion
Equations", J. Comp. Phys. 160:214-282.

\bibitem{KL00}
A.~Kurganov and Doron Levy (2000), "A Third-Order Semidiscrete
Central Scheme for Conservation Laws and Convection-Diffusion
Equations", SIAM J.~Sci.~Comput.~22:1461-1488.
\bibitem{MoL}
O.~A.~Liskovets (1965), Differential equations I 1308-1323.

\bibitem{SSP}
C.-W.~Shu and S.~Osher (1988), "Efficient implementation of
essentially non-oscillatory shock-capturing schemes",
J.~Comp.~Phys.~v77:439-471.

\bibitem{LO96}
Xu-dong Liu and S.~Osher (1996), "Nonoscillatory High Order
Accurate Self-Similar Maximum Principle Satisfying Shock Capturing
Schemes I", SIAM J.~Numer.~Anal.~33:439-471.

\bibitem{Sod}
G.~A.~Sod (1978), "A Survey of Several Finite Difference Methods
for Systems of Nonlinear Hyperbolic Conservation Laws".
J.~Comput.~Phys.~27:1-31.

\bibitem{Lax_test}
P.~D.~Lax (1954), "Weak solutions of nonlinear hyperbolic
equations and their numerical computation", Comm.~Pure
Appl.~Math.~7:159–193.

\bibitem{OT}
S.~A.~Orszag and C.~M.~Tang (1979), "Small-scale structure of
two-dimensional magnetohydrodynamic turbulence". J.~Fluid
Mech.~90, 1:129-143.

\bibitem{MC}
B.~Van Leer (1977), "Towards the ultimate conservative difference
scheme III. Upstream-centered finite-difference schemes for ideal
compressible flow". J.~Comp.~Phys.~23: 263-75.

\bibitem{Torr}
M.~Torrilhon (2003), "Non-uniform convergence of finite volume
schemes for Riemann problems of ideal magnetohydrodynamics".
J.~Comput.~Phys.~192: 73-74.

\bibitem{WoodCol84}
P.~Woodward and P.~Colella (1984), "The numerical simulation of
two-dimensional fluid flow with strong shock".
J.~Comput.~Phys.~54: 115–173.

\bibitem{SZS03}
J.~Shi, Y-T.~Zhan, and C-W.~Shu (2003), "Resolution of high-order
WENO schemes for complicated flow structures",
J.~Comput.~Phys.~186: 690–696.

\bibitem{WZL04}
Z.~J.~Wang, L.~Zhang and Y.~Liu (2004), "Spectral finite volume
method for conservation laws on unstructured grids IV: extension
to two-dimensional systems", J.~Comput.~Phys.~194: 716–741.

\bibitem{KO}
B.~Gustafsson, H.~O.~Kreiss and J.~Oliger (1995), "Time Dependent
Problems and Difference Methods". Wiley-Interscience (New York).




\end{thebibliography}
\end{document}